\documentclass[review, 3p]{elsarticle}

\usepackage{amssymb}

\usepackage[dvipsnames]{xcolor}
\usepackage{import}
\usepackage{comment}
\usepackage{booktabs} 
\usepackage{multirow}
\usepackage{pifont}  
\usepackage{setspace}
\usepackage{amsmath}
\usepackage{amssymb}
\usepackage{pifont}
\usepackage{graphicx}
\usepackage{subcaption}
\usepackage{hyperref}
\usepackage{cleveref}
\usepackage{todonotes}
\journal{Computer Speech and Language}

\begin{document}

\begin{frontmatter}



\title{To train or not to train adversarially: A study of bias mitigation strategies for speaker recognition}

\author{Raghuveer Peri}
\author{Krishna Somandepalli}
\author{Shrikanth Narayanan}

\address{Department of Electrical and Computer Engineering, University of Southern California, Los Angeles, CA, USA\\
\{rperi, somandep\}@usc.edu, shri@ee.usc.edu }


\begin{abstract}
Speaker recognition is increasingly used in several everyday applications including smart speakers, customer care centers  and other speech-driven analytics. 
It is crucial to accurately evaluate and mitigate biases present in machine learning (ML) based speech technologies, such as speaker recognition, to ensure their inclusive adoption.
ML fairness studies with respect to various demographic factors in modern speaker recognition systems are lagging compared to other human-centered applications such as face recognition.
Existing studies on fairness in speaker recognition systems are largely limited to evaluating  biases at specific operating points of the systems, which can lead to false expectations of fairness. 
Moreover, there are only a handful of bias mitigation strategies developed for speaker recognition systems.
In this paper, we systematically evaluate the biases present in speaker recognition systems with respect to gender across a range of system operating points. 
We also propose adversarial and multi-task learning techniques to improve the fairness of these systems. 
We show through quantitative and qualitative evaluations that the proposed methods improve the fairness of ASV systems over baseline methods trained using data balancing techniques.
We also present a \textit{fairness-utility} trade-off analysis to jointly examine  fairness and the overall system performance. 
We show that although systems trained using adversarial techniques improve fairness, they are prone to reduced utility. On the other hand, multi-task methods can improve the fairness while retaining the utility. These findings can inform the choice of bias mitigation strategies in the field of speaker recognition.
\end{abstract}




\end{frontmatter}


\begin{spacing}{1}
\section{Introduction}
\label{sec:intro}
Consider a home security system that authenticates the homeowner based on their voice - what if it works reliably only for individuals from certain demographic groups? 
``What is the practical applicability of such a system''?, ``Is this system \textit{fair}?'', ``How do we identify biases in this system?'', and ``How might we mitigate these biases?''.
Such questions are being addressed at a rapid pace in technology domains such as computer vision and natural language understanding that primarily rely on machine learning (ML) algorithms --- leading to the emergence of \textit{ML fairness} as a field of study in its own right \cite{barocas-hardt-narayanan}. 
In the context of speech technologies, ML fairness studies are mostly limited to applications such as speech-to-text conversion~\cite{koenecke2020racial}.
Very few studies have considered ML fairness for speaker recognition which is a key component in applications such as personalized speech technologies and voice-based biometric authentication. 

Speaker recognition is the task of identifying a person based on their voice.
Automatic speaker verification (ASV), which is a specific application of speaker recognition, refers to the task of authenticating users based on their voice characteristics. It has found widespread adoption in smart home appliances (e.g., Alexa) \cite{asv_article}, voice authentication in airports \cite{cornacchia2020user} and, as a biometric system in customer service centers and banks \cite{asv_nuance, asv_citi}.
With the proliferation of speech technologies in everyday lives, biases present in these systems can have dire social consequences.
There is an imminent need to identify, understand and mitigate biases in these systems. Our goal in this work is to systematically evaluate biases in speaker recognition, and study bias mitigation strategies. Specifically, we examine whether adversarial or multi-task learning techniques help mitigate these biases, and improve the fairness of speaker recognition systems. We present experimental findings that demonstrate the conditions under which fairness can be improved, and 
 have made the related code and model information publicly available for the benefit of the research community\footnote{Code and information about pre-trained models can be found in \url{https://github.com/rperi/trustworthy-asv-fairness}}.

Generally, ASV applications have different expectations of performance depending on the end use-case. For example, security applications typically impose strict restrictions on the proportion of \textit{impostors}\footnote{A person attempting to maliciously gain access to a biometric system claiming to be a different person} they erroneously admit. expect fewer instances of incorrectly rejecting the user's voice.
On the other hand, smart home applications may value user convenience i.e., they value seamless verification of \textit{genuine} users at the expense of tolerating more impostor cases.
Thus, it is crucial to consider biases that arise in different applications where ASV systems are deployed. While tremendous strides have been made in evaluating and improving fairness in applied ML fields \cite{barocas-hardt-narayanan}, techniques for bias mitigation in ASV systems are limited \cite{fenu2020improving, shenimproving}.
Notably, most current bias evaluation frameworks for ASV are restricted to specific operating points of the systems \cite{toussaint2021sveva, fenu2020exploring}, limiting their applicability to specific use-cases.
In particular, differences in the equal error rate~(EER) of an ASV system between different demographic population groups is commonly used as a proxy for the biases present in that system \cite{fenu2020exploring, shenimproving}. EER is the error of the system where the rate of accepting impostors is equal to the rate of rejecting genuine users. It refers to a specific operating point of the ASV system, and fairness evaluations using differences in EER may not generalize to other use-cases.
More general conclusions about the fairness of the systems (applicable to distinct ASV use-cases) require a thorough evaluation of biases at several system operating points.
In addition, \textit{utility} of ASV systems, which can be understood as the overall performance (not considering any specific demographic group) is an important consideration in evaluating the practical applicability of bias mitigation strategies. An ideal bias mitigation strategy is expected to reduce the differences in performance between the different demographic groups, with minimal degradation of the utility of the system.

Bias mitigation strategies in ML systems often involve models trained using data balancing~\cite{serna2021insidebias, zhang2020towards, zhao2018gender}.
In the context of fairness in ASV systems, data balancing methods were studied with respect to age and gender \cite{fenu2020improving}. However, it is not evident if such techniques are the most suitable to induce fairness. For example, in other ML fields such as computer vision, studies have shown that data balancing may not be sufficient to mitigate biases (e.g,~\cite{wang2019balanced}).
Another class of techniques to improve fairness tackle biases in the modeling stage by incorporating fairness constraints during training \cite{pmlr-v28-zemel13, zafar2017fairness}.
When demographic information (e.g., gender, language background, age etc.) is available, it can be used in an \textit{adversarial training}~(AT) setup to learn \textit{speaker embeddings}~(compact speech representations that capture information about speaker's identity) that are fair with respect to the demographic attributes \cite{li2021estimating}.
Adversarial methods for ASV systems typically train encoders to learn speaker-discriminative representations while stripping them of demographic information \cite{noe2020adversarial}. 
However, these \textit{demographic factors} are considered as components of a person's identity \cite{hassan2021soft}, and can help improve the ASV system performance \cite{luu2020leveraging}.
For example, it is typically easier to reject an impostor verification claim, when the impostor's gender is different from that of the target speaker. Removing gender-related information from speaker embeddings using adversarial techniques can lead to degraded ASV performance, reducing the utility of such systems.
It may be beneficial to develop ASV systems that perform equally well for people belonging to different demographic groups, despite the speaker embeddings retaining information of the demographic attributes.

For this purpose, a \textit{multi-task} learning~(MTL) strategy can be employed to simultaneously predict factors related to speaker identity such as gender, age or language. Techniques leveraging demographic information have been shown to achieve improved ASV performance, where models are trained to predict speaker labels along with age and nationality \cite{luu2020leveraging}. 
In a recent work on fairness in ASV, Shen et al. \cite{shenimproving} developed a system to fuse scores from separately trained gender-specific models. However, this method does not explicitly infuse demographic information into the speaker embeddings, but merely combines the separate gender-specific scores. To the best of our knowledge, there exists no current work on multi-task training techniques that leverage demographic information to train speaker embeddings with the goal of improving fairness.

In a recent work~\cite{peri2020robust}, we showed that adapting unsupervised adversarial invariance (\textbf{UAI})~\cite{jaiswal2018unsupervised} --- an adversarial method proposed for computer vision tasks ---  for speaker representation learning makes the ASV system robust to adverse acoustic conditions. 
In the current work, we extend this framework by proposing adversarial training (\textbf{UAI-AT}) and multi-task learning (\textbf{UAI-MTL}) for bias mitigation in ASV systems that use pre-trained neural speaker embeddings.
The various bias mitigation strategies we study for ASV are summarized in Table~\ref{tab:prior}, alongside some exemplar studies in automatic speech recognition (ASR), which is a speech technology field, and other ML domains. 
Our experiments not only evaluate the fairness of the proposed UAI-AT and UAI-MTL methods but also compare them with data balancing, UAI, AT and MTL methods as baselines.
In addition to extensive fairness analyses, we jointly examine the overall performance of the ASV system. This is referred to as \textit{fairness-utility} trade-off.
\newline Our specific contributions and findings are summarized below:\newline
\begin{table}[]
\centering
\caption{Summary of research works leveraging different techniques to improve fairness of deep learning models. The column labeled "Other ML domains" lists a few exemplar works dealing with fairness in fields other than speech technologies. There is limited research in tackling biases in speech technologies compared to non-speech domains notably in methods for improving fairness of ASV systems. We propose to employ adversarial (AT,UAI-AT) and multi-task (MTL,UAI-MTL) techniques to improve the fairness of ASV systems.}
\label{tab:prior}
\resizebox{0.7\textwidth}{!}{
\begin{tabular}{c|c|c|c}
\toprule & ASV   & ASR  & Other ML domains \\ \midrule
Data balancing     & \cite{fenu2020improving}                    & \cite{sari2021counterfactually}, \cite{garnerin2021investigating} & \begin{tabular}[c]{@{}c@{}}\cite{zhao2018gender}, \cite{klare2012face}, \cite{yu2021fair}\end{tabular}  \\ \midrule
AT             &                                \ding{55} & \cite{tripathi2018adversarial}, \cite{yadav2021masknet}, \cite{sun2018domain}  & \cite{wang2019balanced}, \cite{ravfogel2020null}, \cite{han2021diverse}, \cite{alvi2018turning}, \cite{zhang2018mitigating}, \cite{jaiswal2020invariant}                                                                                                   \\
MTL          &           \cite{shenimproving}       &      \ding{55}     & \cite{Das_2018_ECCV_Workshops}, \cite{ryu2017inclusivefacenet}, \cite{oneto2019taking},\cite{vaidya2020empirical}                                                      \\ 
UAI-AT       &                         Ours                         &          \ding{55}     &             \cite{jaiswal2019unified}                                                                                                  \\
UAI-MTL      &          Ours    &          \ding{55}      &         \ding{55}                                              \\ \bottomrule
\end{tabular}
}
\end{table}
\textbf{Summary of Contributions:}
\begin{itemize}
    \item We systematically evaluate biases present in ASV systems at multiple operating points. Current work in this field is mostly focused on a single operating point, which can lead to an incomplete evaluation of fairness. To this end, we adopt the fairness discrepancy rate metric \cite{de2020fairness} to measure the fairness of ASV systems under different system operating conditions. 
    \item
    We propose novel adversarial and multi-task training methods to improve the fairness of ASV systems. As shown in Table \ref{tab:prior}, adversarial and multi-task techniques for bias mitigation in ASV systems are limited or non-existent.
    We compare the proposed methods against baselines that rely on data balancing, using quantitative and qualitative evaluations.
    \item 
    In addition to fairness evaluations, we also consider the utility, which is the overall system performance, using standard performance metrics such as EER. Joint considerations of fairness and utility can help inform the choice of bias mitigation techniques.
\end{itemize}
\textbf{Summary of findings}:
\begin{itemize}
    \item We show that the fairness of baseline ASV systems (trained using data balancing) with respect to gender varies with the operating point of interest. Our experiments show increased bias of the baseline methods as the system operation moves to regions with fewer instances of incorrectly rejecting genuine users. We demonstrate that, compared with the baseline systems, the fairness of the proposed adversarial and multi-task methods have minimal dependence on the operating point.
    \item 
    We demonstrate using qualitative visualizations and quantitative metrics that the proposed techniques are able to mitigate biases to a large extent compared to the baseline systems based on data balancing. We further show that this observation holds true across a range of different operating conditions.
    \item We observe that the adversarial technique improves fairness but suffers with reduced utility. In contrast, the multi-task technique is able to improve fairness while retaining the overall system utility. These findings can inform choosing appropriate bias mitigation strategies, while carefully considering the target use-case of the speaker recognition systems.
\end{itemize}

The rest of the paper is organized as follows.
In section \ref{sec:lit}, we provide background of existing work related to fairness in ASV. Section \ref{sec:method} details the methodology used to induce fairness in ASV systems, followed by a description of the metrics we use to evaluate the fairness of ASV systems in Section \ref{sec:metrics}. We provide a brief description of the datasets used to build and evaluate our models in Section \ref{sec:dataset}. Section \ref{sec:experiment} outlines the baselines and experiments designed to investigate the biases in the developed ASV systems (including ablation studies). This is followed by the corresponding results and discussions in Section \ref{sec:results}. Finally, we conclude the paper in Section \ref{sec:conclusion}, where we summarize our findings, and provide avenues for potential future research. \\ \\
\textbf{A note on fairness terminology as it relates to this work \\}
Extensive research has been done on fairness in diverse domains such as law, social science, computer science, philosophy etc. There exist equally diverse definitions of fairness, each given within the context of that particular domain \cite{mulligan2019thing}. With the recent proliferation of ML algorithms in socio-technical systems, fairness in ML has garnered immense interest. The general idea of \textit{the absence of any prejudice or favoritism toward an individual or a group based on their inherent or acquired characteristics} has been termed \textit{fairness} in the ML literature \cite{mehrabi2019survey}, and the field of study dealing with these issues is referred to as Fair-ML \cite{barocas-hardt-narayanan}.
Most work evaluating and tackling fairness in ML systems can be broadly categorized into two: \textit{group fairness} and \textit{individual fairness} \cite{binns2020apparent}. Group fairness deals with ensuring that the outcomes~(correct predictions or errors) of a classifier are equally distributed between different demographic groups \cite{chouldechova2017fair}.
On the other hand, individual fairness requires that two individuals who are similar to each other be treated similarly \cite{dwork2012fairness}. 
In the context of this work, we focus on group fairness, leaving individual fairness for future work.
\section{Background}
\label{sec:lit}

The goal of an ASV system is to automatically detect whether a given speech utterance belongs to a claimant who is a previously \textit{enrolled} speaker.
ASV techniques in the past represented speakers using statistical methods leveraging gaussian mixture models. Likelihood ratio tests were then used to determine if a speech utterance belonged to an enrolled speaker \cite{reynolds1995automatic}.
More recently, embedding based methods have been developed where the speech utterances are modeled using low-dimensional, speaker-discriminative representations~\cite{ivector-basic, snyder2018x}.
In particular, embedding-based techniques that model a speaker's vocal characteristics using deep-learning methods have gained prominence \cite{snyder2018x, variani2014deep}.

In this section, we first provide a brief overview of a typical deep-learning based ASV system. Next, we present studies that examine fairness in ASV systems, vis-a-vis other ML application domains. In particular, we discuss related-work of two key aspects relevant to this paper: (1) Evaluation of biases in ASV systems (2) Bias mitigation strategies, which include adversarial and multi-task methodologies.
\begin{figure*}[t!]
\centering
\includegraphics[width=0.95\textwidth]{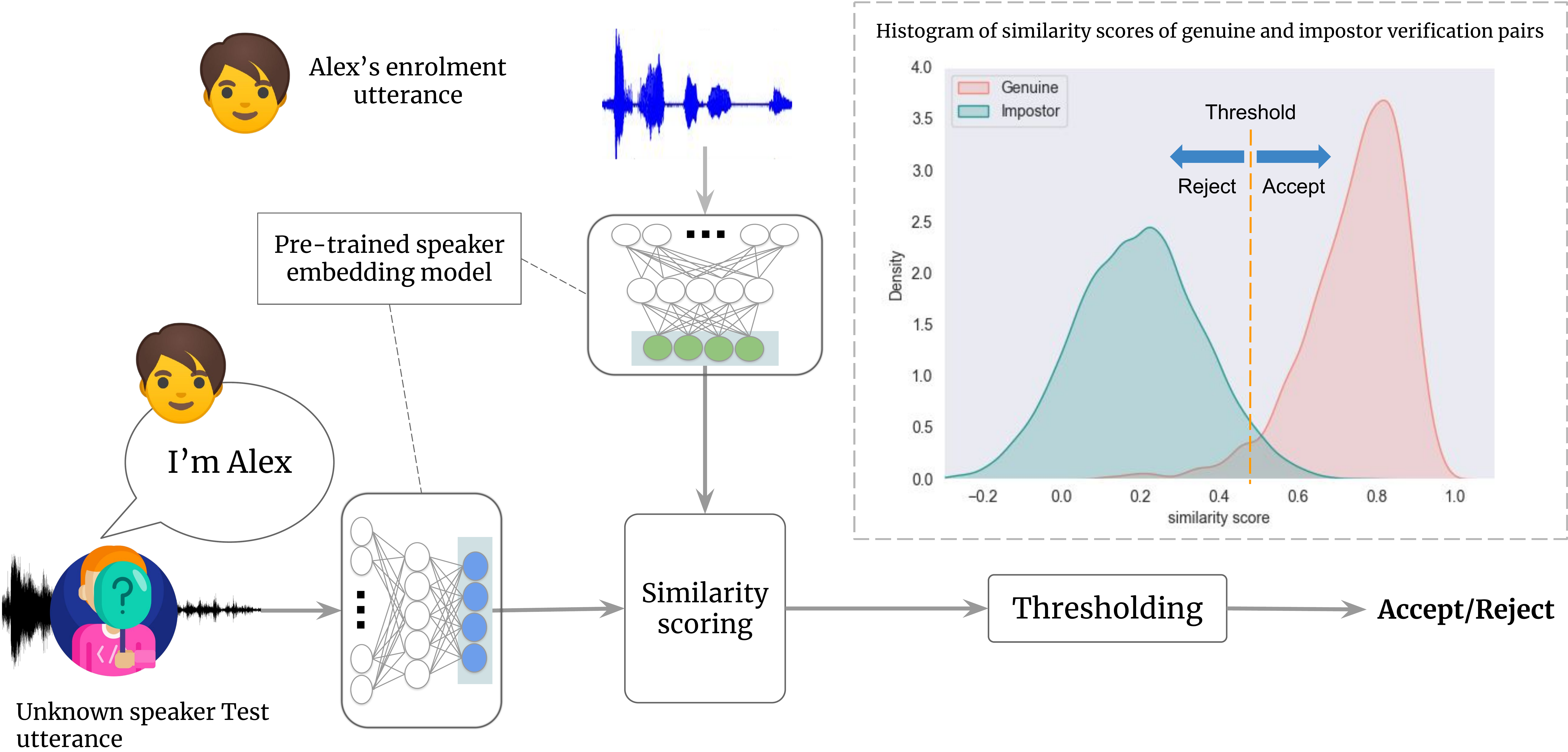}
\caption{Block diagram of a typical deep-learning based ASV system. Pre-trained speaker embedding models are used to represent the enrolment speech (from a pre-enrolled speaker) and the test speech (uttered by the user who is attempting to be verified). Similarity between the representations is computed and a threshold is applied on the similarity scores. The verification claim is accepted if the similarity score exceeds the threshold. The inset figure shows a histogram of the similarity scores of a typical ASV system. Genuine verification pairs with scores smaller than the threshold contribute to false rejects, while impostor verification scores greater than the threshold contribute to false accepts. The threshold determines the operating point, and can be chosen suitable to the use-case. An overview of ASV is provided in Section \ref{ssec:lit_asv}.}
\label{fig:asv}
\vspace{-0.15in}
\end{figure*}
\subsection{Overview of ASV}
\label{ssec:lit_asv}
As shown in Figure \ref{fig:asv}, speaker embeddings from the test utterance and a previously collected enrolment utterance are obtained using pre-trained speaker embedding models. The speaker embedding models are typically trained in a fully-supervised setting on large amounts of speech with speaker identity labels using a deep neural network \cite{snyder2018x}. The goal is to learn an \textit{embedding} of the speech utterance that is discriminative of the speaker. Much like the likelihood ratio tests, the similarity between the embeddings of the enrolment and test utterances is compared to a threshold to \textit{verify} the identity of the speaker in the considered speech utterance. 

Verification pairs consisting of speech from the same speaker in both the enrolment and test utterances are called \textit{genuine} pairs. When the speaker of the test utterance is different from the enrolment utterance, the corresponding verification pairs are called \textit{impostor} pairs. 
The inset plot in Figure \ref{fig:asv} shows the histogram of similarity scores of genuine and impostor verification scores of a typical ASV system. 
Since the final output of an ASV system is a binary accept/reject decision, its errors maybe classified into two categories: false accepts (FA) and false rejects (FR). FAs are instances where impostor pairs are incorrectly accepted, while FRs are instances where genuine pairs are incorrectly rejected by the ASV system. There exists a trade-off between the number of FAs and FRs of the system depending on the threshold on the similarity scores. Systems which use a larger threshold tend to have fewer FAs and more FRs. This can be useful in high-security applications such as in border security. Similarly, a smaller threshold can be used to reduce the number of FRs in applications requiring greater user convenience. Thus, the chosen similarity threshold determines the \textit{operating point} of the overall ASV system, and it can be tuned to suit the end application. Equal error rate~(EER) is a commonly used metric to evaluate performance of ASV systems. It captures the performance at a single operating point, where the false acceptance rate~(FAR) is equal to the false rejection rate~(FRR).

As noted in \cite{stoll2011finding} and \cite{si2021exploring}, the verification scores of pairs comprising speakers from the female and male demographic groups can be significantly different. Furthermore, as shown in Figure \ref{fig:asv}, the decision by the ASV system depends on the similarity threshold of these verification scores. Therefore, any biases present in the verification scores propagate to the final decisions of the ASV system. This can result in unfair treatment of certain demographic groups. 

\subsection{Fairness in ASV}
Issues of fairness and biases have been extensively studied in several domains involving ML. A few prominent examples include facial analysis \cite{buolamwini2018gender, klare2012face, robinson2020face, 250171, howard2018ugly, beveridge2009factors, howard2019effect, ryu2017inclusivefacenet}, natural language understanding \cite{sap2019risk, bolukbasi2016man, park2018reducing, diaz2018addressing, dixon2018measuring}, affect recognition \cite{xu2020investigating, stoychev2022effect, gorrostieta2019gender}, criminal justice \cite{green2018fair, chouldechova2017fair, wadsworth2018achieving, 10.1145/3442188.3445902}, and health care \cite{10.1145/3442188.3445917, 10.1145/3442188.3445934}.

In the context of speech technology research, the intersection of speech ML and fairness is mostly limited to automatic speech recognition~(ASR). For example, Koenecke et al. \cite{koenecke2020racial} found racial disparities in the performance of several state-of-the-art commercial ASR systems. Feng et al. \cite{feng2021quantifying} have analyzed the biases in a Dutch ASR system with respect to gender, age etc. 
Evaluations of ASR systems using criterion commonly used in Fair-ML research have been explored extensively \cite{liu2021modelbased, 9610166, liu2021measuring, garnerin2021investigating}. 
However, a systematic evaluation of fairness in ASV systems is scarce in current literature. 

A recent work explored racial and gender disparities in several speaker classification systems \cite{chen2022exploring}. However, their analysis was restricted to closed-set classification task, which is different from the more challenging ASV setup we consider in this work.
The field of biometrics is perhaps the most relevant in the context of speaker verification from the perspective of an end application.
As noted by Drozdowski et al. \cite{drozdowski2020demographic}, a majority of bias detection and mitigation works in biometrics focus on face recognition \cite{250171, beveridge2009factors, ryu2017inclusivefacenet, klare2012face}, and some in fingerprint matching \cite{galbally2018study, preciozzi2020fingerprint}. Fairness in voice-based biometrics remains to be an under-explored field with only a handful of works \cite{fenu2020improving, toussaint2021sveva, fenu2021fair, shenimproving}.

\subsubsection{Evaluating biases in ASV systems}
\label{ssec:lit_faireval}
As discussed in Section, \ref{ssec:lit_asv}, ASV systems can be tailored for different applications depending on the threshold used to accept/reject verification pairs.
Studies that evaluate biases in ASV systems need to ensure that conclusions drawn from them are not limited to specific applications. Evaluations of biases that only focus on specific operating points of an ASV system can lead to incomplete conclusions about their fairness.
Some studies in the past have documented differences in the performance of ASV systems owing to demographic attributes such as gender and age \cite{stoll2011finding}, where ASV systems that use statistical speaker models were analyzed. However, the analysis was restricted to verification scores alone, while the biases in final system decision were not considered. Furthermore, it is unclear if the findings translate to contemporary  speaker-embedding based methods. In a recent work, Si et al. \cite{si2021exploring} showed significant differences in the speaker verification scores between certain demographic groups, but it is not evident how such differences affect an ASV system in practice. 
Owing to the recent challenges organized by NIST \cite{sadjadi20172016, Sadjadi2019}, there has been an increased focus on improving ASV performance across languages \cite{torres2017ll,lee2019i4u}. Significant improvements were obtained on the evaluation data provided in the challenge. Multi-lingual analyses were added in the latest VoxSRC challenge \cite{brown2022voxsrc}. However, performance evaluations in these studies used only a specific operating point characterized by the EER. 

In a more recent work, fairness of ASV systems with respect to age and gender as the demographic factors has been explored \cite{fenu2020exploring}. However, the evaluation of fairness was again limited to disparity in EER between the demographic groups. Toussaint et al. \cite{toussaint2021sveva} proposed an evaluation framework for probing the fairness of ASV systems. Their evaluations focused on the minimum detection cost function, which again considers one particular operating point characterized by a threshold on the speaker verification scores. They provide visualizations of the verification scores and the detection error trade-off (DET) curves, which offer a qualitative way of analyzing the biases at several operating points.
It might be of interest to quantify fairness at different operating points to systematically understand how the system behaves. Such an analysis can be critical to understanding the behavior of the ASV systems in different types of applications, for e.g., high security applications requiring low FAR (fewer incorrectly accepted impostor pairs) and high-convenience applications requiring low FRR (fewer incorrectly rejected genuine pairs).

Fenu et al. \cite{fenu2021fair} performed fairness evaluations using several different definitions of fairness at multiple operating points with a focus on data balancing strategies to mitigate biases.
In this work, we adopt the fairness discrepancy rate ~(FaDR) metric that has been proposed recently in the biometrics literature \cite{de2020fairness} to evaluate fairness.
FaDR computed at different system operating points can be used to systematically analyze the fairness of the proposed methods. A detailed explanation of the FaDR metric can be found in Section \ref{sec:metrics}.\\ \\
\textbf{Fairness-Utility trade-off:\ }
Studies have shown that humans demonstrate a superior speaker identification performance in their native language compared to non-native languages \cite{perrachione2007learning}. Similarly, ASV systems can be biased to perform better for certain demographic populations \cite{fenu2020exploring}.
Although reliance on demographic information can potentially improve ASV performance, it can lead to unfair systems that discriminate against certain demographic groups. Thus, there is a trade-off between fairness and utility. Such trade-offs have been studied extensively in general Fair-ML literature \cite{zhao2019inherent, calders2009building, haas2020price, du2020fairness}. For example, Zhao et al. \cite{zhao2019inherent} have shown that adversarial training methods to improve fairness reduce the utility of such systems. However, empirical studies demonstrating such trade-offs between fairness and utility in ASV systems are limited \cite{fenu2021fair}. In this work, we study how the proposed techniques perform in improving fairness, while also evaluating their utility using standard metrics.


\subsubsection{Mitigating biases in ASV systems}
\label{ssec:lit_fairimprove}
Prior work has demonstrated differences in the performance of ASV systems across gender groups \cite{reynolds2000speaker}.
This led to the development of gender-specific models that were used in combination with a gender classification module in ASV \cite{bimbot2004tutorial, kanervisto2017effects}. This was the case even in the popular i-vector based ASV models \cite{ivector-basic}.
However, such a methodology of training separate models for each demographic group needs the demographic group label (either self-reported by the speaker or predicted by a model) at the time of inference. This information may not always be available, or possible to infer.
In addition, such methods can further perpetuate biases and undermine certain privacy criteria by requiring the systems to infer demographic attributes.
Therefore, for practical purposes, it is desirable to develop unified, demographic-agnostic ASV models. Most of the recent deep learning based approaches train a unified model agnostic to the demographic groups, while trying to ensure substantial representation from each group in the training data \cite{chung2018voxceleb2}. However, such systems can still be prone to issues of biases because they are not explicitly trained to induce fairness.

In fair-ML literature, algorithms to improve fairness or mitigate biases fall into one of the three categories: \textit{pre-processing}, \textit{post-processing} and \textit{in-processing} \cite{mehrabi2019survey}.
A common pre-processing method to develop fair models is by training them using data that is balanced with respect to the various potential sources of bias \cite{zhang2020towards}. This approach has been explored in ASV systems, where data from individuals that is balanced with respect to genders, languages, and ages is used to train models to improve fairness \cite{fenu2020improving}. Post-processing techniques are used when only access to a pre-trained model is available, and when it is impractical to train models using our own data \cite{bellamy2018ai}. However, such techniques are commonly employed in closed-set classification tasks, and it is not straightforward to generalize them to a verification setup like ASV. In-processing techniques involve explicitly inducing fairness into the model during training by introducing fairness constraints \cite{berk2017convex}. A common method is adversarial techniques that use demographic information during training to learn \textit{de-biased} representations \cite{zhang2018mitigating}. When demographic labels are available, they can also be used in a multi-task fashion. In such methods, the demographic labels are used to reduce the performance disparity between groups \cite{xu2020investigating}.\\ \\
\textbf{Adversarial Training and Multi-task Learning:\ }
 In the adversarial training methodology, labels are used to learn representations devoid of the demographic information. 
 Since the representations contain very little demographic information, the systems are likely to use information discriminative of the primary prediction task, and rely less on demographic attributes, making the systems fair with respect to those attributes \cite{zhang2018mitigating}. 
 Adversarial training methods have shown promise in developing unbiased ML-based classification systems \cite{edwards2015censoring, zhang2018mitigating}. 
 Some studies in face recognition (e.g., \cite{morales2020sensitivenets}) have shown the efficacy of adversarial techniques to improve fairness. However, it is not evident if those benefits translate to a verification setup (ASV systems in particular). 
 Language invariant speaker representations for ASV were developed using adversarial methods \cite{bhattacharya2019adapting}. However, evaluations were limited to a single operating point characterized by EER. No{\'e} et al. \cite{noe2020adversarial} employ adversarial training techniques to develop speaker identity representations that are devoid of certain sensitive attributes. However, their goal was privacy protection and not to improve fairness. To the best of our knowledge, such techniques have not been used for developing fair ASV systems, and we explore that direction in this work.
 
 Demographic labels could also be used in a multi-task methodology with the label prediction as a secondary task to learn \textit{demographic-aware} representations. This can be particularly useful in tasks dealing with detecting identity such as face and speaker recognition, where the demographic attributes form part of the person's identity. In such cases, instead of stripping the representations of demographic factors, one can train models to ensure that the performance of the systems is similar across demographic groups. 
 Xu et al. \cite{xu2020investigating} propose to provide attribute information to a facial expression recognition model in an attribute-aware fashion to improve its fairness. However, these observations were made on a classification task, which is different from our verification setup. 
 In biometric settings (which is closer to our target ASV task), multi-task training methods can be used to add demographic information to the general-purpose representations. 
Luu et al. \cite{luu2020leveraging} have shown that demographic attribute information can be used in a multi-task setup to improve utility of ASV systems. However, fairness of such systems trained using the multi-task training setup is not studied. 
In a more recent work, Shen et al. \cite{shenimproving} have shown that gender-specific adaptation of encoders to extract separate gender-specific representations can improve the fairness of ASV systems. This can be treated as a demographic-aware method, and they show that their method can improve the fairness while also improving the overall utility. 
However, fairness evaluations were limited to differences in EER between the genders.
We intend to investigate if using a multi-task setup to train demographic-aware speaker representations can improve the ASV utility in addition to reducing the differences in performance between the different demographic groups.
\section{Methods}
\label{sec:method}
We develop methods to transform existing speaker embeddings to another representation space with the goal of minimizing biases with respect to demographic groups in ASV systems. This is achieved by training models using demographic labels in addition to the speaker identity labels. 
We explore adversarial and multi-task techniques to train the embedding models to improve the fairness of pre-trained speaker embeddings.
\begin{figure*}[t!]
\centering
\includegraphics[width=0.7\textwidth]{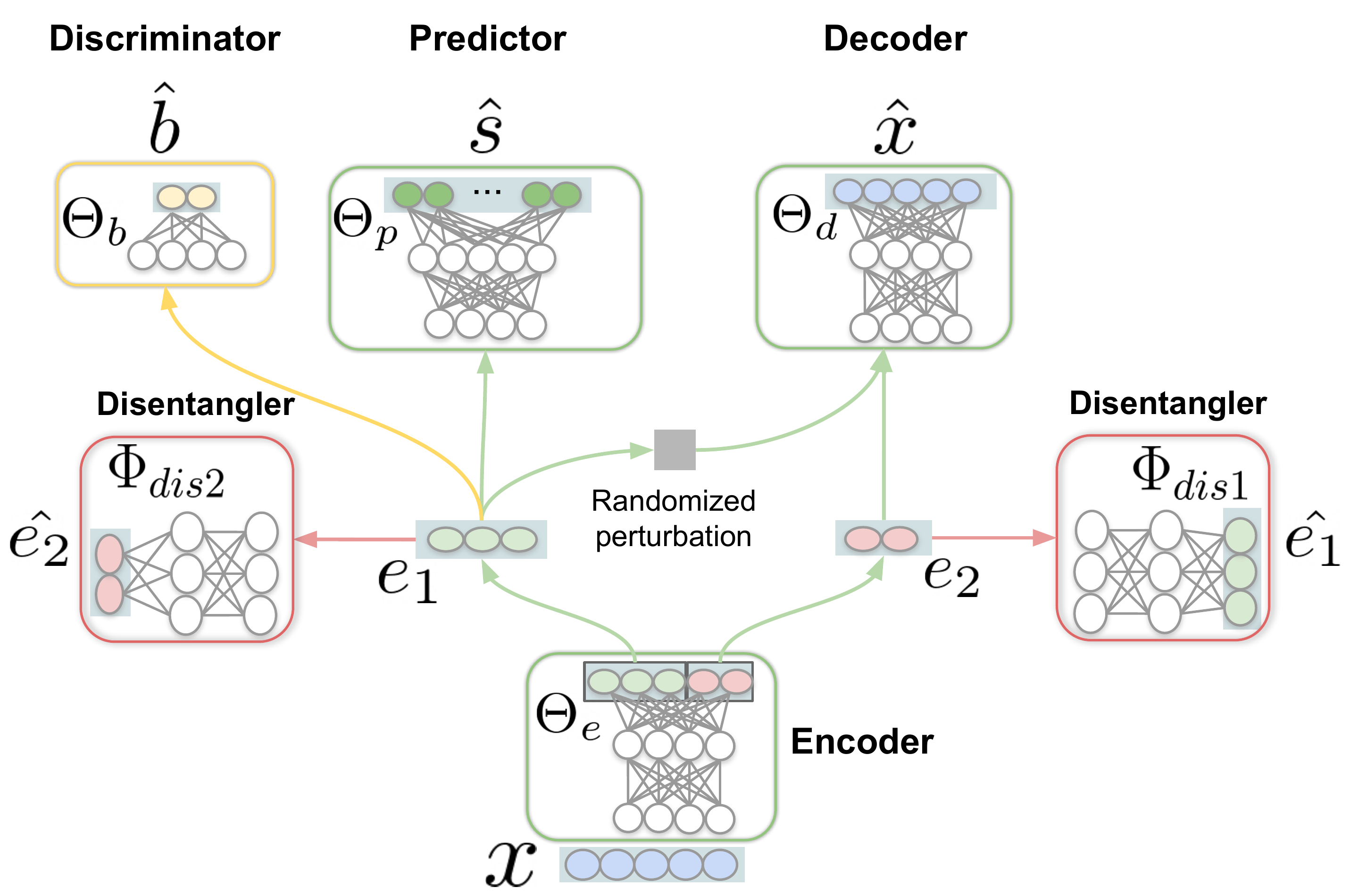}
\caption{Block diagram of the method showing the predictor and decoder modules to predict speaker labels and reconstruct the original input respectively, and the disentanglers to reconstruct the different embeddings from one another. The primary branch (consisting of modules shown in green bounding boxes) is pitted against the secondary branch (shown by modules in red bounding boxes) goal is to train the encoder to learn nuisance-invariant speaker representations in $e_1$ and all other information in $e_2$. The discriminator module (yellow bounding box) is tasked with predicting the demographic factor (e.g., gender) from $e_1$, and can be trained in an adversarial setup (UAI-AT) or in a multi-task (UAI-MTL) setup  with the predictor. The adversarial setup would learn demographic-invariant speaker embeddings, while the multi-task setup would learn demographic-aware speaker embeddings in $e_1$.}
\label{fig:unifai}
\vspace{-0.15in}
\end{figure*}
We employ the unsupervised adversarial invariance~(UAI) framework, which was originally proposed in \cite{jaiswal2018unsupervised}. We had adapted this approach in our previous work to disentangle speaker factors from \textit{nuisance factors} unrelated to the speaker's identity present in x-vectors \cite{periempirical}. However, as noted by Jaiswal et al. \cite{jaiswal2019unified}, the UAI technique cannot be directly used to induce invariance to demographic factors in an unsupervised fashion, and demographic labels are needed to induce invariance using adversarial techniques. Therefore, we propose to use the adversarial extension of the UAI technique developed by Jaiswal et al. \cite{jaiswal2019unified}. In addition, we also propose a novel multi-task extension to the UAI framework. Figure \ref{fig:unifai} shows the schematic diagram of the proposed method. The techniques including UAI and its adversarial and multi-task extensions are explained in detail below.

\subsection{Unsupervised adversarial invariance (UAI)}
\label{ssec:UAI}
The central idea behind this technique is to project the input speaker representations into a split representation consisting of two embeddings, referred to as $\mathbf{e}_{1}$ and $\mathbf{e}_{2}$ in Figure \ref{fig:unifai}. While $\mathbf{e}_{1}$ is trained with the objective of capturing speaker-specific information, $\mathbf{e}_{2}$ is trained to capture all other nuisance factors. This is achieved by training two branches in an adversarial fashion. 

\begin{equation}
L_{prim} = \alpha L_{pred}\left(\mathbf{s},\hat{\mathbf{s}}\right) + \beta L_{recon}\left(\mathbf{x},\hat{\mathbf{x}}\right) 
\label{eq:L_main}
\end{equation}

\begin{equation}
L_{sec} = L_{dis1}(\mathbf{e}_{2},\hat{\mathbf{e}}_{2}) + L_{dis2}(\mathbf{e}_{1},\hat{\mathbf{e}}_{1}) 
\label{eq:L_adv}
\end{equation}

\begin{equation}
\label{eq:L_all}
\begin{aligned}
    \underset{\Theta_{prim}}{\textnormal{min}} \hspace{1pt} \underset{\Phi_{sec}}{\textnormal{max}} \; L_{prim} & + \gamma L_{sec}, \\
    \textnormal{where} \; \Theta_{prim} = {\Theta_{e}} \cup {\Theta_{d}} \cup {\Theta_{p}}, & \; \Phi_{sec} = \Phi_{dis1} \cup \Phi_{dis2}
\end{aligned}
\end{equation}

The goal of one branch, called \textit{primary} branch (consisting of the encoder, predictor and decoder shown in green bounding boxes in Figure \ref{fig:unifai}), is to predict speakers using $\mathbf{e}_{1}$ as input (using the \textit{predictor} module) and reconstruct the x-vectors using $\mathbf{e}_{2}$ and a randomly perturbed version of $\mathbf{e}_{1}$ as input (using the \textit{decoder} module). The random perturbation ensures that the network learns to treat $\mathbf{e}_{1}$ as an unreliable source of information for the reconstruction task, hence forcing $\mathbf{e}_{1}$ to not contain information about factors other than the speaker. The perturbation of $\mathbf{e}_{1}$ is modelled as a dropout module that randomly removes some dimensions from $\mathbf{e}_{1}$ to create a noisy version denoted by $\mathbf{e}_{1}^\prime$. The primary branch produces the loss term shown in Equation \ref{eq:L_main}, where $L_{pred}$ is modelled as categorical cross-entropy loss to predict speakers, and $L_{recon}$ is modelled as mean squared error (MSE) reconstruction loss of the decoder. The terms $\Theta_{e}, \Theta_{d}, \Theta_{p}$ denote the network parameters of the encoder, decoder and predictor respectively as shown in Figure \ref{fig:unifai}. The speaker prediction task forces $\mathbf{e}_{1}$ to capture speaker-related information, while the reconstruction task ensures that $\mathbf{e}_{2}$ captures information related to all factors.

The other branch, called the \textit{secondary} branch (consisting of the disentanglers shown in red bounding boxes in Figure \ref{fig:unifai}), is trained to minimize the mutual information between $\mathbf{e}_{1}$ and $\mathbf{e}_{2}$. This is achieved in the \textit{disentangler} module consisting of two networks that predict $\mathbf{e}_{1}$ from $\mathbf{e}_{2}$ and vice-versa. 
The secondary branch produces the loss term given in Equation \ref{eq:L_adv} which is the sum of the two disentangler losses, each of which is modelled as MSE loss. The terms $\Phi_{dis1}, \Phi_{dis2}$ denote the network parameters of the disentangler modules shown in Figure \ref{fig:unifai}. The UAI model is trained with a minimax objective shown in Equation \ref{eq:L_all} by alternating between the primary and secondary branch updates according to a fixed schedule. The parameters $\alpha, \beta, \gamma$ control the contribution of the prediction, reconstruction and the disentanglement loss terms respectively. Detailed explanation of the method to disentangle speaker representations can be found in \cite{peri2020robust}.

As reported in our previous work \cite{periempirical}, a characteristic of this technique is that it disentangles the speaker identity from nuisance factors, which are all the factors unrelated to the speaker's identity \cite{periempirical}, such as acoustic noise, reverberation etc.
Jaiswal et al. \cite{jaiswal2019unified} proposed an extension to the UAI technique, called unified adversarial invariance, which uses the demographic labels to induce invariance to those attributes. We explore this framework, in addition to a multi-task extension to improve fairness of ASV systems.

\subsection{Adversarial and multi-task extensions of UAI: UAI-AT and UAI-MTL}
\label{sssec:uaiat_uaimtl}
UAI by itself cannot provide invariance to demographic factors. Therefore, Jaiswal et al. \cite{jaiswal2019unified} extended the UAI framework to include demographic labels during training. In particular, they introduced a \textit{discriminator} that is used to predict demographic labels. In the formulation proposed in \cite{jaiswal2019unified}, this discriminator (shown in yellow bounding box in Figure \ref{fig:unifai}) is trained in an adversarial fashion along with the disentangler of the UAI. 
\begin{equation}
\label{eq:uai-param}
\begin{aligned}
& \underset{\Theta_{prim}}{\textnormal{min}} \hspace{1pt} \underset{\Phi_{sec}}{\textnormal{max}} \; L_{prim} + \gamma L_{sec} + \delta L_{bias}({b},\hat{{b}}) \\
\textnormal{UAI-AT:} \quad \Theta_{prim} & = {\Theta_{e}} \cup {\Theta_{d}} \cup {\Theta_{p}}, \quad \Phi_{sec} = \Phi_{dis1} \cup \Phi_{dis2} \cup \Theta_{b} \\
\textnormal{UAI-MTL:} \quad \Theta_{prim} & = {\Theta_{e}} \cup {\Theta_{d}} \cup {\Theta_{p}}  \cup \Theta_{b}, \quad \Phi_{sec} = \Phi_{dis1} \cup \Phi_{dis2}
\end{aligned}
\end{equation}
Equation \ref{eq:uai-param} shows the training objective that includes the discriminator loss denoted by $L_{bias}$, which is modeled as cross-entropy loss between the true and predicted demographic labels denoted as $b$ and $\hat{b}$ respectively. We denote the method of adversarially training the discriminator along with UAI as UAI-AT throughout the rest of the paper. The term corresponding to UAI-AT in Equation \ref{eq:uai-param} shows how the discriminator (with set of trainable parameters denoted by $\Theta_{b}$ being part of the secondary branch) is trained adversarially with the predictor. This ensures that the learned embeddings $\mathbf{e}_{1}$ do not retain demographic information, thereby achieving the desired invariance. The discriminator loss is modelled as categorical cross-entropy loss between the true and predicted demographic labels.

On the other hand, it is not evident if adversarial training to induce invariance to demographic factors is necessary to learn fair representations. Given the demographic labels, they can be used to train the discriminator in a multi-task (as opposed to adversarial) fashion. We call these demographic-aware speaker representations, and this method is denoted as UAI-MTL in the rest of the paper.
The term corresponding to UAI-MTL in Equation \ref{eq:uai-param} shows how the discriminator parameters (being part of the primary branch) are trained in a multi-task fashion with the predictor. The objective is to learn a representation that captures speaker identity information while retaining the demographic attribute information. In both the UAI-AT and UAI-MTL methods, the parameter $\delta$ controls the contribution of the discriminator loss to the overall loss term.
\section{Metrics}
\label{sec:metrics}
In this section, we provide details of the metrics that we use to evaluate the fairness and utility of ASV systems. A brief description of each metric is also provided in Table \ref{tab:metric} for a quick reference. 

\subsection{Utility: Equal error rate (EER)}
EER refers to a particular operating point of the system where the FAR equals FRR. This metric is commonly used to evaluate the utility of ASV systems. Lower values of EER signify better system utility. We chose EER over the minimum detection cost function (minDCF), which is another commonly used evaluation metric, as minDCF requires specifying parameters such as the relative costs of the detection errors and the target speaker prior probability, which imply a particular application \cite{van2007introduction}. We wanted to avoid introducing additional variability arising due to the different parameters. Note that we only use EER to measure utility and not to evaluate fairness.

\subsection{Fairness: Fairness discrepancy rate (FaDR)}
There exist several metrics to measure and evaluate fairness of ML systems, some of which are more suited to a particular application than others. Garg et al. \cite{garg2020fairness} have discussed several commonly used metrics proposed in the fairness literature. Metrics such as equal opportunity and equalized odds have been extensively studied. Verma and Rubin \cite{verma2018fairness} discuss how some metrics can deem an algorithm fair which the other metrics have deemed unfair. Therefore, it is crucial to choose a metric that satisfies the notion of fairness we aim to achieve. As discussed in Section \ref{sec:lit}, a reasonable goal of fairness in ASV systems is to ensure that the performance differences between demographic groups is small across a range of different operating points. Algorithms that are fair only at certain operating points can result in a false sense of fairness, and can be detrimental when used to design systems with real-world impact.

A straight-forward way to analyze the fairness of biometric systems is to use the disparity in EER between the demographic groups (termed as differential outcomes by Howard et al. \cite{howard2019effect}) as an indication of the fairness. This method has been previously used in evaluating fairness of ASV systems \cite{fenu2020improving, shenimproving}. However, this approach assumes that each demographic group has its own threshold on the verification scores. This can lead to false notions of fairness, because in most real-world systems a single threshold is used for verification irrespective of the demographic group \cite{de2020fairness}.
In order to overcome this limitation, Periera and Mercel \cite{de2020fairness} propose a metric called fairness discrepancy rate~(FaDR) to account for FARs and FRRs in biometric systems. They propose to evaluate fairness at multiple thresholds that can be chosen agnostic of the demographic groups. We employ this metric to evaluate the fairness of our models.
\begin{equation}
\label{eq:FaDR}
\begin{aligned}
    FaDR(\tau) = 1 - & (\omega A(\tau) + (1-\omega) B(\tau)) \\
    A(\tau) = |FAR^{g_1}(\tau) - FAR^{g_2}(\tau)|,& \ \ B(\tau) = |FRR^{g_1}(\tau) - FRR^{g_2}(\tau)|
\end{aligned}
\end{equation}
Intuitively, FaDR computes the weighted combination of absolute differences in FARs and FRRs between demographic groups. The threshold $\tau$ is applied on demographic-agnostic verification trials to compute the demographic-agnostic FAR (corresponding to the zero-effort score distribution used by Periera and Mercel \cite{de2020fairness}), which characterizes a particular operating point of the system. The fairness of a system can be measured at different values of the threshold $\tau$ corresponding to different operating points. Assuming two demographic groups are of interest, at a given threshold $\tau$, FaDR\footnote{This definition is a special case of FaDR when only two demographic groups are present. A more general definition can be found in \cite{de2020fairness}, } is defined in Equation \ref{eq:FaDR}
where $FAR^{g_1}(\tau)$ and $FRR^{g_1}(\tau)$ refer to the FAR and FRR, when the threshold is applied on the similarity scores of verification pairs consisting only of speakers belonging to demographic group $g_1$ (similarly for demographic group $g_2$). To contextualize it with the terminology used by Gother et al. \cite{250171}, this can be viewed as a weighted combination of FA and FR differentials, with the \textit{error discrepancy weight} given by $\omega$ ($0<=\omega<=1$).
\begin{table}[]
\centering
\caption{List of metrics used in this paper with a brief description and their purpose (utility or fairness). The FaDR metric evaluates the fairness of the ASV system at a particular operating threshold (characterized by demographic-agnostic FAR), while the area under the FaDR-FAR curve summarizes the fairness at the various operating points of interest with a single number. For the error rates (ranging from 0.0 to 1.0) used to measure utility, lower is better. For the metrics used to measure fairness, higher is better. FaDR values range from 0.0 to 1.0, while the maximum value of auFaDR-FAR depends on the FAR values over which area is computed.}
\label{tab:metric}
\resizebox{0.9\textwidth}{!}{
\begin{tabular}{|c|c|c|c|}
\hline
Metric                    & Abbreviation & Brief description                                                                                                                                       & Purpose  \\ \hline
False Acceptance Rate     & FAR          & Rate of accepting impostor verification trials                                                                                                          & Utility  \\ \hline
False Rejection Rate      & FRR          & Rate of rejecting genuine verification trials                                                                                                           & Utility  \\ \hline
Demographic-agnostic FAR  & -            & \begin{tabular}[c]{@{}c@{}}FAR computed on demographic-agnostic\\ verification trials\end{tabular}                                                      & Utility  \\ \hline
Equal Error Rate          & EER          & \begin{tabular}[c]{@{}c@{}}Error rate corresponding to threshold \\ where FAR equals FRR\end{tabular}                                                   & Utility  \\ \hline
Fairness Discrepancy Rate & FaDR         & \begin{tabular}[c]{@{}c@{}}Weighted absolute discrepancy in FAR and FRR \\ between demographic groups (Equation \ref{eq:FaDR})\end{tabular} & Fairness \\ \hline
Area under FaDR-FAR curve & auFaDR-FAR   & \begin{tabular}[c]{@{}c@{}}Area under the FaDR curve plotted \\ at several thresholds\end{tabular}                                                      & Fairness \\ \hline
\end{tabular}
}
\end{table}
\subsubsection{Note on error discrepancy weight ($\omega$)} FaDR can be computed by weighing the discrepancy between the demographic groups in $2$ different types of errors, FAR and FRR. The error discrepancy weight, $\omega$ in Equation \ref{eq:FaDR}, can be used to determine the importance of the different types of errors. $\omega=1.0$ corresponds to the case where the differences between the demographic groups are evaluated only using their FARs. Similarly, $\omega=0.0$ corresponds to considering the differences only in the FRRs between demographic groups. $\omega=0.5$ reflects the condition that discrepancy between the demographic groups in FAs and FRs are equally important. Intuitively, it can be used to weigh the relative importance of the discrepancy in FAR and FAR between the demographic groups. For example, evaluating FaDR at high values of $\omega$ could be useful in applications such as in border control where FAs are critical \cite{de2020fairness}. A larger emphasis can be given to reducing demographic disparity in accepting impostor verification pairs. Similarly, smaller values of $\omega$ can be used to evaluate the fairness in applications such as in smart speakers where considering FRRs that can degrade the user experience is more important.

\subsection{Fairness: Area under the FaDR-FAR curve (auFaDR-FAR)} FaDR can be computed at various operating points of an ASV system by varying the threshold on verification similarity scores. These thresholds are applied on demographic-agnostic verification scores to compute demographic-agnostic FARs. Therefore, we can obtain a curve showing the FaDR of the system at various demographic-agnostic FAR values, and this curve can be used to compare the fairness of different systems. Furthermore, Pereira and Mercel \cite{de2020fairness} propose the use of area under the FaDR-FAR (auFaDR-FAR) curve as an objective summary of the fairness of a system for various conditions. We use this as the primary metric for evaluation because it summarizes the fairness of systems at the operating points of interest.


\section{Dataset}
\label{sec:dataset}

In this section, we provide details of the datasets used for training and evaluating our models. We employed different subsets of the Mozilla Common Voice~(MCV) dataset \cite{commonvoice:2020} in our experiments. In addition, we also used a subset of the Voxceleb1 dataset as an out-of-domain evaluation set. The MCV corpus consists of crowd-sourced audio recordings of read speech collected from around the world in multiple languages. It is an ongoing project, where any user with access to a computer or smart phone and an internet connection can upload speech samples for research purposes. Users are prompted to read sentences appearing on their screen, and these recordings are validated by other users. We also used the Voxceleb1 dataset \cite{nagrani2017voxceleb} as an external corpus (different from the MCV corpus) to evaluate the generalizability of the described methods on out-of-domain data. It consists of \textit{in-the-wild} recordings of celebrity interviews with speaker identity labels. Unlike in the MCV corpus, the gender labels in Voxceleb1 were not self-reported but obtained from Wikipedia. The subsets of these corpora we use in our experiments are described below, and their statistics are provided in Table \ref{tab:data_classif}. 
\subsection{Training}
\label{sec:data_train}
We use the following datasets to train the speaker embedding transformation model using the methods described in Section \ref{sec:method}. These datasets consist of speech samples with speaker identity labels and demographic labels such as gender and language.
\begin{itemize}
    \item xvector-train-U~(`U' stands for `Unbalanced'): A subset of MCV employed to train the model (described in Section \ref{ssec:expt_baseline}) that was used to extract the baseline speaker embeddings. It corresponds to the data referred to as \textit{Train-2} condition in \cite{fenu2020improving}. This subset consists of recordings in English and Spanish, which are not balanced with respect to gender, as shown in Table \ref{tab:data_classif}. 
    
    \item xvector-train-B~(`B' stands for `Balanced'): Another subset of MCV employed to train the x-vector model used to extract the baseline speaker embeddings. It corresponds to data referred to as \textit{Train-1} condition in \cite{fenu2020improving}. This subset consists of recordings which are balanced with respect to the number of speakers per gender and age, as shown in Table \ref{tab:data_classif}. This is a subset of the xvector-train-U data. 
    
    \item embed-train: Data used to train the proposed models to improve fairness using the methods described in Section \ref{sec:method}. Pre-trained speaker embeddings extracted on this data were used to train our models. This is a subset of the xvector-train-U data.
    
    \item embed-val: This subset was created with the same speakers present in embed-train dataset to tune the parameters of the models by evaluating speaker and demographic prediction on data unseen during training. The speaker and demographic prediction accuracies on this subset can be treated as a proxy for the amount of information in the intermediate speaker embeddings pertaining to speaker identity and demographic factors respectively.
\end{itemize}
\begin{table}[]
\centering
\caption{Statistics of datasets used to train and evaluate speaker embedding models. xvector-train-B is balanced with respect to gender in the number of speakers, while xvector-train-U is not balanced. embed-train and embed-val (used to train proposed models) have different utterances from the same set of speakers to facilitate evaluating speaker classification performance during embedding training. eval-dev and eval-test (used to evaluate ASV utility and fairness) have speech utterances with no overlap between speakers. voxceleb-H is an out-domain evaluation dataset, and unlike all the other datasets, is not collected from the MCV corpus. \textbf{\#spk.}-number of unique speakers, \textbf{\#samples}-number of speech utterances in training or number of verification pairs in evaluation, \textbf{F}-Female, \textbf{M}-Male}
\label{tab:data_classif}
\resizebox{0.95\textwidth}{!}{
\begin{tabular}{cc|cc|cc|ccc}
\toprule
                          &        & xvector-train-U & xvector-train-B & embed-train & embed-val & eval-dev & eval-test & voxceleb-H \\ \midrule
\multirow{2}{*}{\textbf{\#spk.}} & F & 664 & 620        & 585     &      585 & 1194        & 529        & 527 \\ 
& M   & 1706      & 620       &  1692         &     1692       & 5231        & 2231       & 665    \\
                          \midrule
\multirow{2}{*}{\textbf{\#samples}} & F &  86,332    & 87,949    & 51,016  &      12,989 & 721,370         & 545,103 & 226,690 \\ 
& M   & 124,179   & 101,527   &  117,918 &    30,205 & 633,126         & 528,666 & 324,206  \\
                          \bottomrule
\end{tabular}
}
\end{table}

\subsection{Evaluation}
\label{sec:data_eval}
The following datasets are used to evaluate the transformed speaker embeddings for their utility and fairness in ASV. 
\begin{itemize}
    \item eval-dev: We use this data to create development set verification pairs to fine-tune hyperparameters of our models, such as the bias weight in Equation \ref{eq:uai-param}. The speakers in this subset are unseen during training (speakers not present in any of the subsets described in Section \ref{sec:data_train}). Tuning hyperparameters on this subset using metrics useful for verification allows us to build models that are better suited for the task of speaker verification. Roughly 1.3M verification pairs were created from this data. Evaluations were performed on separate subsets of the pairs corresponding to different genders. For example, to evaluate verification performance on the female demographic group, pairs were created using enrolment and test utterances only from speakers belonging to the female demographic group.
    
    \item eval-test: 
    Similar to eval-dev data described above, this contains recordings from speakers not present in any of the above datasets. Particularly, there is no speaker overlap with the eval-dev dataset. Verification pairs from this data are used to evaluate models in terms of both fairness and utility. This dataset was used as held-out data to evaluate only the best models (after hyperparameter tuning). More than 1M verification pairs were created from this data.
    
    \item voxceleb-H: Following Toussaint et al. \cite{toussaint2021sveva}, we performed evaluations on the voxceleb-H split. It is a subset of Voxceleb1 containing $1190$ speakers, and ~500K verification trials consisting of same gender and same nationality pairs. Different from the MCV corpus which is mostly read speech, the Voxceleb1 dataset consists of recordings from celebrity interviews in an unconstrained setting. This dataset facilitates fairness evaluations of ASV systems in more relaxed settings consisting of spontaneous speech.
\end{itemize}

\section{Experiments}
\label{sec:experiment}
In Section \ref{sec:method}, we described methods to transform pre-trained speaker embeddings to induce fairness. In this section, we describe the experiments designed to evaluate the fairness and utility of the proposed UAI-AT and UAI-MTL methods, by comparing them against suitable baselines. In addition, we describe the ablation studies we performed to investigate the importance of the different modules used in our methods. \\ \\
\textbf{Setup}: Our method consists of training an embedding transformation model using speaker identity and demographic labels in a closed-set classification setup. For this paper, we restrict our analyses to gender\footnote{We use the term \textit{gender} to refer to the self-reported gender in the datasets, except for Voxceleb, where the labels were obtained from Wikipedia. We restrict our analysis to binary gender categories due to the limitation imposed by the availability of labels in existing speech datasets \cite{garnerin2021investigating}, and hope to overcome these limitations in the future.} as the demographic attribute for which fairness is desired, but the proposed methods can be extended to other demographic attributes (e.g., age) as well.
The encoder from the trained speaker representation model is used to extract embeddings, that are then evaluated for fairness and utility in a speaker verification setting. Below we describe the baselines along with the training setup of the proposed methods. We then discuss the evaluation setup and implementation details.
\subsection{Baselines}
\label{ssec:expt_baseline}
The pre-trained speaker embeddings used as input to our models were chosen from the prior methods developed to improve fairness in ASV systems \cite{fenu2020improving}. We compare our methods against ASV systems developed using these chosen off-the-shelf embeddings as baselines, which allows us to investigate the effectiveness of our proposed methods in improving the fairness of existing speaker embeddings.
\begin{itemize}
    \item x-vector-U: As a weak baseline, we use the pre-trained models\footnote{\url{https://drive.google.com/drive/folders/1FW7FqkNuw2QqsaZ6PVF7EzLLg2ZjKbQ7}} that were trained using data not balanced with respect to gender. Specifically, the models were trained using the \textit{xvector-train-U} dataset described in Section \ref{sec:dataset}. Evaluation on this baseline provides an understanding of the biases present in speaker verification systems trained using unbalanced data. This is particularly important because most existing speaker verification systems rely on speaker embedding models trained on large amounts of data, typically without specific attention to data balancing.
    
    \item x-vector-B: Data balancing is a common technique used to develop fair ML systems. Fenu et al. \cite{fenu2020improving} have employed this strategy to improve fairness of speaker verification systems. This is a stronger baseline against which the proposed UAI-AT and UAI-MTL methods are compared. We use pre-trained models\footnote{\url{https://drive.google.com/drive/folders/1sGq0WO9pw7P6VQXy6ovm64kidu7ue5dE}} that were trained using the \textit{xvector-train-B} dataset described in Section \ref{sec:dataset}.
\end{itemize}
\subsection{Proposed methods}
We trained models with the following methods using gender labels along with the speaker labels on the embed-train dataset described in Section \ref{sec:data_train}. As mentioned before, the embed-train dataset is a subset of the xvector-train-U dataset (Though, in theory we could use the full xvector-train-U dataset, we were able to obtain only a subset due to missing recordings). In contrast with the xvector-train-B dataset, the training data samples are not balanced with respect to the gender labels. The advantage of the proposed methods are that they can leverage all the available data without explicit data balancing.

We used the speaker embeddings referred to as x-vector-B in the previous section \ref{ssec:expt_baseline} as input to our models. The rationale behind using these embeddings was that these were trained using an existing data-balancing technique and have shown to improve fairness \cite{fenu2020improving}. This allowed us to explore the proposed techniques as a means to further improve the fairness of existing ASV systems that are already trained to reduce biases.
\begin{itemize}
    \item UAI-AT: As described in Section \ref{sssec:uaiat_uaimtl}, the gender labels can be used in addition to the UAI technique, similar to the technique proposed by Jaiswal et al. \cite{jaiswal2019unified} to improve fairness. As shown in Table \ref{tab:uai_check}, all modules including the discriminator from Figure \ref{fig:unifai} were employed. The optimization was implemented as an alternating mini-max game, where the predictor training forces the encoder to retain speaker information, while the discriminator training forces it to strip demographic information. In the minimization step the encoder and the predictor from Figure \ref{fig:unifai} were updated while keeping the secondary branch (discriminator and disentanglers) frozen for a few iterations. In the maximization step, the encoder and the secondary branch were updated keeping the primary branch frozen. This way, the encoder was trained to retain speaker identity information while discarding information about the demographic attributes from the intermediate speaker representations. In practice, instead of maximizing the discriminator loss, we minimized the loss between the predictions and a random sampling of the gender labels from the empirical distribution obtained from the training data, similar to the technique used in \cite{jaiswal2019unified}, \cite{alvi2018turning}.
    \item UAI-MTL: Different from the adversarial training strategy, here the gender labels were used in a multi-task fashion. Similar to the UAI-AT technique, predictor training forces the encoder to retain speaker information. However, in this case, the discriminator is trained in a multi-task fashion using gender labels to explicitly force the encoder to learn demographic information, producing demographic-aware speaker embeddings. This is achieved by making the discriminator a part of the primary branch, with the secondary branch consisting of only the disentanglers.
\end{itemize}
\begin{table}[t!]
\centering
\caption{Table showing active blocks (corresponding to Figure \ref{fig:unifai}) used in different embedding transformation techniques. All the techniques use an encoder to reduce the dimensions of the input speaker embeddings and a predictor to classify speakers. The NLDR and UAI techniques do not require a disentangler since they do not use demographic attribute labels to achieve invariance. The MTL and AT techniques do not employ the disentangler module. The UAI-AT and UAI-MTL techniques use discriminator with demographic attribute labels in addition to the disentangler module. The first four rows denote ablation experiments, while the last two correspond to the proposed techniques.}
\label{tab:uai_check}
\resizebox{0.75\textwidth}{!}{
\begin{tabular}{cccccc}
\toprule
        & Encoder   & Predictor & Decoder   & Disentangler & Discriminator \\ \midrule
NLDR     & \checkmark & \checkmark &  &    &      \\
UAI     & \checkmark & \checkmark & \checkmark & \checkmark    &     \\
MTL     & \checkmark & \checkmark &  &    & \checkmark     \\
AT      & \checkmark & \checkmark &  &    & \checkmark     \\ \midrule
UAI-AT  & \checkmark & \checkmark & \checkmark & \checkmark    & \checkmark     \\ 
UAI-MTL & \checkmark & \checkmark & \checkmark & \checkmark    & \checkmark     \\ \bottomrule
\end{tabular}
}
\end{table}
\subsection{Ablation studies}
\label{ssec:expt_ablation}
As discussed in Section \ref{sec:method} and shown in Table \ref{tab:uai_check}, the proposed UAI-AT and UAI-MTL techniques use all the modules including the encoder, predictor, decoder, disentanglers and discriminators. We performed ablation studies to better understand the impact of each module on the performance by selectively retaining certain modules. We also considered the scenario where gender labels are not available.
In such scenarios, we investigate if fairness can be improved either by UAI or by simple dimensionality reduction using neural networks, which we term non-linear dimensionality reduction (NLDR). This is in contrast to linear dimensionality reduction approaches such as principal component analysis. The modules corresponding to Figure \ref{fig:unifai} that are active in these experiments are shown in Table \ref{tab:uai_check}.
\begin{itemize}
    \item Non-linear dimensionality reduction (NLDR): We investigate the effect of non-linear transformation of speaker embeddings while retaining speaker identity information without considering the demographic information. This is achieved by training a neural network to transform pre-trained speaker embeddings using only the speaker labels. This helps understand if simple dimensionality reduction techniques can provide benefits in terms of reducing the biases in the systems. We denote this experiment as NLDR, since the model is trained using just the encoder and predictor modules with non-linear activation functions. 
    
    \item UAI: As described in Section \ref{sec:method}, the UAI technique was used to improve the robustness of speaker verification systems to nuisance factors such as acoustic noise, reverberation etc., that are not correlated with speaker's identity \cite{peri2020robust}. However, since demographic attributes such as gender and age are related to the speaker's identity, we had observed that this method does not remove these biases from the speaker embeddings \cite{periempirical}. We evaluate if such training can improve the fairness without the need for demographic label information.
    As shown in Table \ref{tab:uai_check}, all modules except the discriminator from Figure \ref{fig:unifai} were used in training.
    
    \item AT: We trained models using the encoder, predictor and discriminator in a standard adversarial setting (without disentanglers). Similar to UAI-AT, the speaker classification task of the predictor forces the encoder to learn speaker-related information, while the discriminator training forces the encoder to learn representations stripped of demographic information.
    The adversarial loss was implemented by training the encoder to minimize predictor loss while maximizing the discriminator loss using alternating minimization and maximization steps. This experiment allowed us to investigate the importance of the disentanglers in the training process.
    
    \item MTL: Similar to the AT setup described above, we used only the discriminator module along with the encoder and predictor modules. In contrast to AT setup, here we trained the discriminator in a multi-task setting with the predictor. This ensured that the encoder retained the speaker information (due to the predictor), and also the demographic information (due to the discriminator). The results of this experiment can be compared with the UAI-MTL method to evaluate the importance of the disentanglers.
    
\end{itemize}
\subsection{Evaluation setup}
\label{ssec:expt_eval}
We used the encoder from the speaker representation models trained using the techniques mentioned above as the embedding transformation module. Specifically, we transformed the x-vector-B speaker embeddings (explained in Section \ref{ssec:expt_baseline}) into a new set of speaker representations using the trained encoders. These transformed speaker embeddings produced from the verification evaluation dataset (Section \ref{sec:data_eval}) were evaluated using the standard ASV setup described in Section \ref{ssec:lit_asv}. We used pre-determined enrollment and test pairs generated from the evaluation data, and compute similarity scores using cosine similarity (inner product of two unit-length vectors). 
We then applied a threshold on the similarity score to produce an accept or reject decision for each verification trial, and the error rates were computed by aggregating the decisions over all the pairs. Varying the threshold produces different error rates, and there exists an inherent trade-off between FAR and FRR. To compute fairness metric (FaDR detailed in Section \ref{sec:metrics}), the FARs and FRRs for each demographic group were separately computed using verification trials belonging to that demographic group. For example, to compute the FAR and FRR for the female population, we aggregated the verification trials where both enrolment and test utterance belonging to the female gender. Following Pereira and Mercel \cite{de2020fairness}, we do not consider cross-gender trials (where enrolment and test utterances belong to different genders), because they tend to produce substantially lower FARs than same-gender trials \cite{hautamaki2010approaching}. To compute the demographic-agnostic FAR values useful for evaluating the auFaDR-FAR metric described in Section \ref{sec:metrics}, we pooled all the verification trials agnostic to their demographic attributes.
\subsubsection{Statistical testing for differences in performance}
\label{sssec:permutation}
We used permutation tests to evaluate the statistical significance of our results. In particular, we used random permutations of the verification scores of the x-vector-B baseline and the proposed methods (UAI-AT or UAI-MTL) to generate a distribution of auFaDR-FAR values. The `true' auFaDR-FPR (without permuting) was compared against this distribution of synthetically generated auFaDR-FAR values to compute the p-value. We used $n=10^4$ permutations on randomly chosen $100,000$ verification trials, with $p<0.01$ to denote significance. For testing the significance of the differences in \%EER, we employed a similar permutation test strategy, but instead used all the verification trials (2M) with $n=10^4$ permutations.
\subsection{Implementation details}
The modules encoder, decoder, predictor and the disentanglers were modeled as multi-layer perceptrons comprising $2$ hidden layers each. The encoder and decoder had $512$ units in each layer, while the disentangler modules had $128$ units in each layer. For the predictor modules, $256$ and $512$ units were used in the first and second hidden layers, respectively. The discriminator module comprised of a $1$ hidden layer network with $64$ hidden units. The probability of dropout used in the random perturbation module was set to $0.75$.

Each model was trained using an early stopping criterion based on the speaker prediction accuracy on the embed-val dataset. In each epoch of the UAI-AT and UAI-MTL training, optimization was performed with $10$ updates of secondary branch for every $1$ update of the primary branch. A minibatch size of $128$ was used, and the primary and secondary objectives were optimized using the Adam optimizer with $1e-3$ and $1e-4$ learning rates, respectively, and a decay factor of $1e-4$ for both. The dimensions for the embeddings $\mathbf{e}_{1}$ and $\mathbf{e}_{2}$ were chosen to be $128$ and $32$, respectively. We set the weights for the losses as $\alpha = 100$, $\beta = 5$ and $\gamma = 100$. The architecture and loss weight parameters were chosen based on our previous work using the UAI technique to improve robustness of speaker embeddings \cite{peri2020robust}. For the discriminator module that is used in the proposed UAI-AT and UAI-MTL methods, we tuned the weight on the bias term denoted by $\delta$ in Equation \ref{eq:uai-param}, by evaluating several models with different weight values on the eval-dev dataset. Table \ref{tab:res_val} in Section \ref{app:bias_weight} shows the fairness (auFaDR-FAR) and utility (\%EER) of systems that were trained with different bias weights on the eval-dev dataset. For each method, the model that gave the best performance (in terms of the auFaDR-FAR on the eval-dev dataset) was used for final evaluations reported in the next section on the held-out eval-test dataset.

\section{Results and Discussion}
\label{sec:results}

In this section, we report results from the experiments described in Section \ref{sec:experiment}, and discuss our findings. 
First, we compare the fairness of the proposed systems against the baselines at a range of system operating points in Section \ref{ssec:res_fair}. We then discuss how these systems compare in terms of their utility in Section \ref{ssec:res_utility}.
Finally, in Section \ref{ssec:res_qualit}, we delve into biases present in the ASV systems at the score level (before the thresholding operation shown in Figure \ref{fig:asv}). This sheds light on the biases present in the verification similarity score distribution of the existing ASV systems, and how the proposed techniques mitigate these biases.
\begin{figure}[t!]
\centering
\begin{subfigure}[b]{.48\linewidth}
\includegraphics[width=\linewidth]{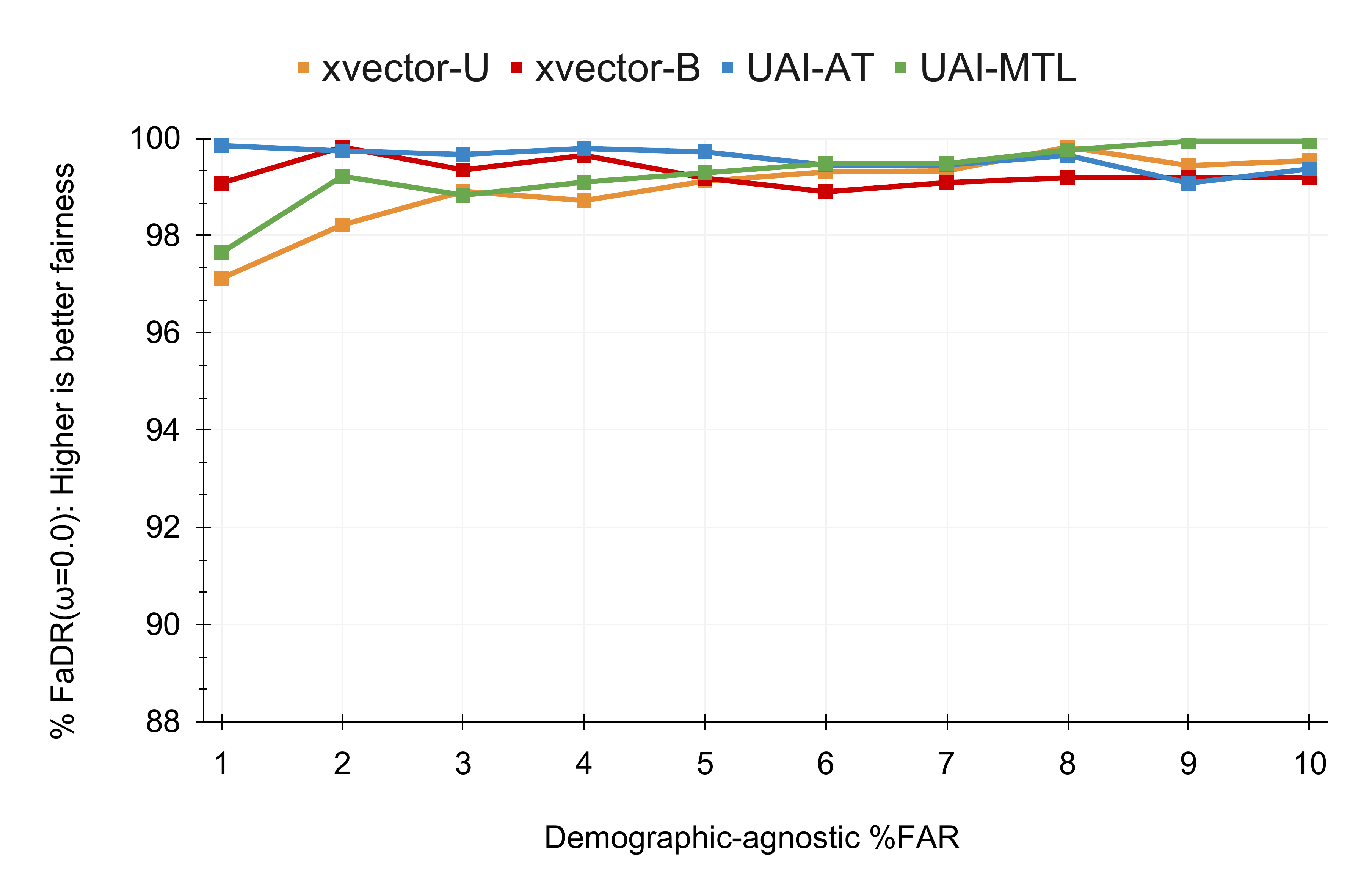}
\caption{\textbf{FaDR with} $\mathbf{\omega=0.0}$. Corresponds to discrepancy in FRR between demographic groups. Useful in applications with emphasis on reducing gender disparity in \textbf{rejecting genuine} verification pairs. Equals $100-(|\text{\%FRR}^{g_1}-\text{\%FRR}^{g_2}|)$}\label{fig:res_test10_0.0}
\end{subfigure}
\begin{subfigure}[b]{.48\linewidth}
\includegraphics[width=\linewidth]{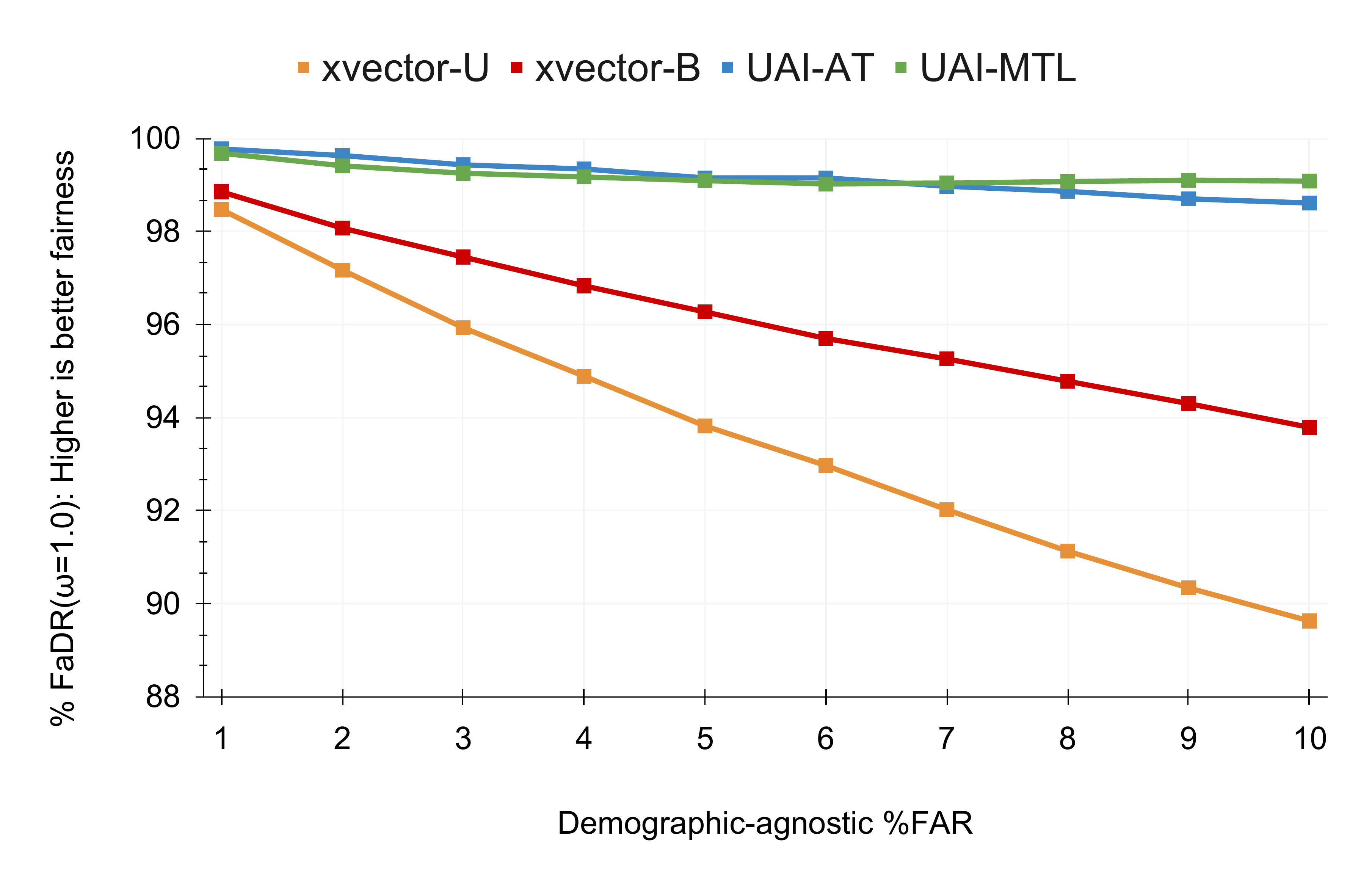}
\caption{\textbf{FaDR with} $\mathbf{\omega=1.0}$. Corresponds to discrepancy in FAR between demographic groups. Useful in applications with emphasis on reducing gender disparity in \textbf{accepting impostor} verification pairs. Equals $100-(|\text{\%FAR}^{g_1}-\text{\%FAR}^{g_2}|)$} \label{fig:res_test10_1.0}
\end{subfigure}

\begin{subfigure}[b]{.60\linewidth}
\includegraphics[width=\linewidth]{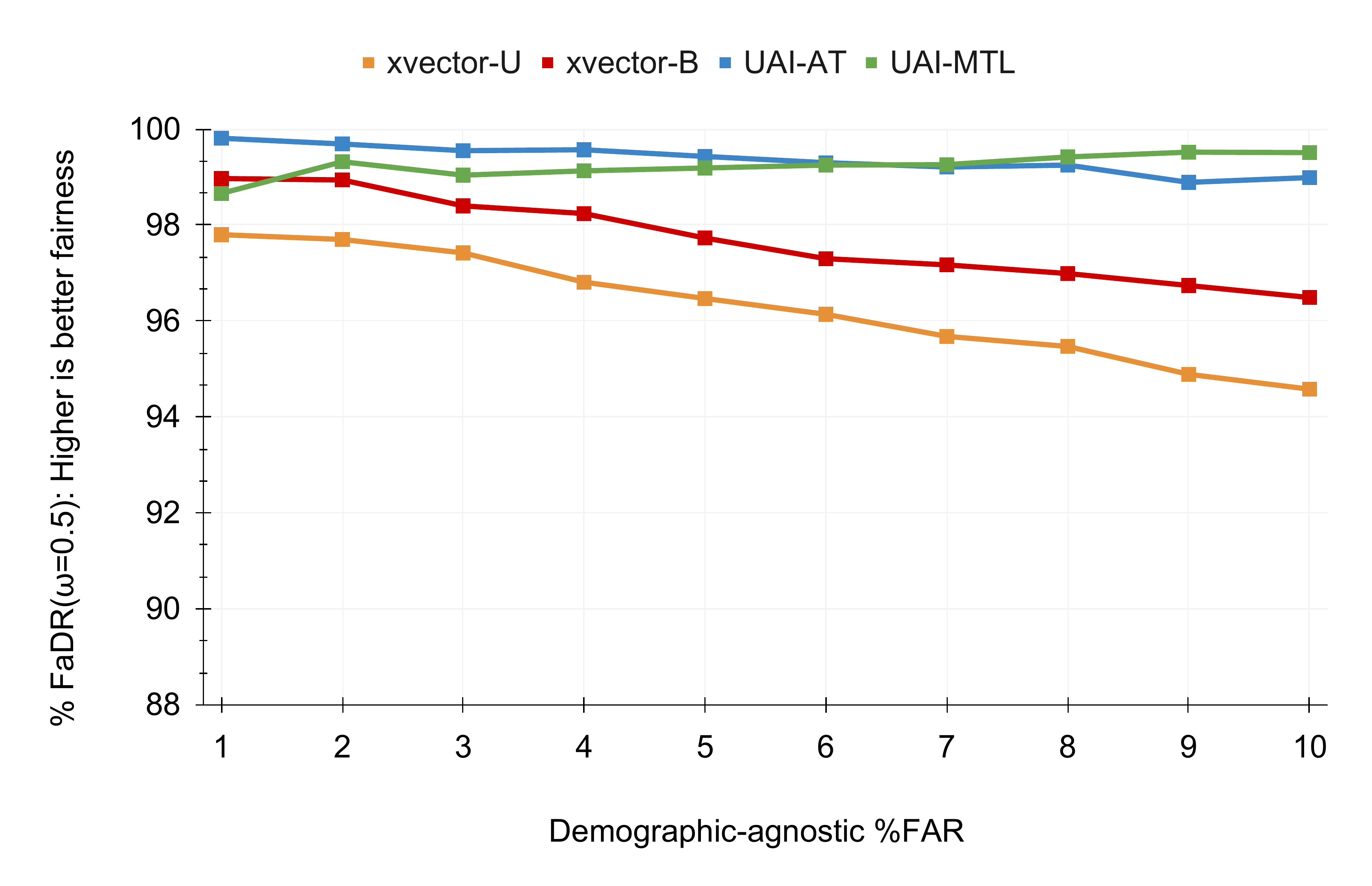}
\caption{\textbf{FaDR with} $\mathbf{\omega=0.5}$. Corresponds to equal contribution of discrepancies in FAR and FRR between genders to the fairness metric. Useful in applications with emphasis on reducing gender disparity both in rejecting genuine verification pairs and accepting impostor verification pairs. Equals $100-(0.5*|\text{\%FRR}^{g_1}-\text{\%FRR}^{g_2}| + 0.5*|\text{\%FAR}^{g_1}-\text{\%FAR}^{g_2}|)$} \label{fig:res_test10_0.5}
\end{subfigure}

\caption{Fairness (binary gender groups) at different operating points characterized by demographic-agnostic FAR upto 10\%, evaluated using $3$ different values for the error discrepancy weight (Eq. \ref{eq:FaDR}), $\omega=0.0, 1.0$ and $0.5$.
Different values of $\omega$ allow evaluating fairness as measured by different error types. 
When evaluating fairness using discrepancy in FRR alone ($\omega=0.0$), there is not much difference between the different systems. All systems seem to perform well with FaDR close to 100\%. When evaluating fairness using discrepancy in FAR alone ($\omega=1.0$), baseline x-vector-B trained on balanced data performs better than x-vector-U. However, the proposed systems (UAI-AT and UAI-MTL) outperform x-vector-B. 
When evaluating fairness using weighted discrepancy in FAR and FRR with equal weights, the proposed systems still show better performance than the baselines. The decision of using techniques to improve fairness of baseline systems is application-specific. Applications requiring a higher emphasis on reduced disparity between demographic groups in accepting impostor authentication claims (higher values of $\omega$) should benefit from the proposed techniques.
}
\label{fig:res_test10}
\end{figure}
\subsection{Fairness}
\label{ssec:res_fair}
Figure \ref{fig:res_test10} shows FaDR plotted at various demographic-agnostic FAR values (upto 10\%) for the proposed UAI-AT and UAI-MTL methods in comparison with the baseline x-vector systems, on the eval-test dataset.
We focus on operating points below 10\% FAR because systems operating at FAR values beyond that may not be useful in practice\footnote{We also performed experiments covering operating points upto 50\% demographic-agnostic FAR, and observed similar trends}. 
The demographic-agnostic FAR values are obtained by applying different thresholds on all verification pairs pooled irrespective of the demographic attribute of the utterances.
FaDR is plotted for $3$ values of the error discrepancy weight~($\omega$ in Eq. \ref{eq:FaDR}), denoting varying amount of contribution from the differences between the genders in FRR and FAR. $\omega=0.0$ corresponds to differences in FRR alone (Figure \ref{fig:res_test10_0.0}), while $\omega=1.0$ corresponds to differences in FAR alone (Figure \ref{fig:res_test10_1.0}). $\omega=0.5$ corresponds to equal contribution of differences in FARs and FRRs (Figure \ref{fig:res_test10_0.5}).

\textbf{Discussion}: From Figure \ref{fig:res_test10_0.0}, we observe that the x-vector systems (red and orange curves) score high on the fairness metric when $\omega=0.0$. This implies that FRR, which is the rate of incorrectly rejecting genuine verification pairs, has minimal dependence on the gender of the speaker. As we discuss later in Section \ref{ssec:res_qualit}, this can be explained from the similarity scores of the x-vector speaker embeddings for the genuine pairs shown in Figure \ref{fig:hist_xvec_bal_genuine}, where we observe a substantial overlap in scores of the female and male populations. Furthermore, we observe that the proposed ASV systems (UAI-AT and UAI-MTL) score similar to the baselines. It can be inferred that if we only care about the FRRs (i.e., how many genuine verification pairs are rejected by the ASV system), then the x-vector systems are already fair with respect to the gender attribute, and additional processing using the proposed methods retains the existing fairness.

On the other hand, as shown in Figure \ref{fig:res_test10_1.0}, the x-vector systems (red and orange curves) are less fair considering the case of $\omega=1.0$. This shows that for the baseline systems, FAR, which is the rate of incorrectly accepting impostor verification pairs, depends on the gender of the speakers. Particularly, the x-vector system trained with imbalanced data scores lower on the fairness metric compared to that trained with balanced data. Furthermore, the fairness of both the x-vector systems drops at higher values of demographic-agnostic FAR. This suggests that data balancing by itself may not achieve the desired fairness at all operating regions of the ASV system considering the biases in FARs between genders. Previous works in domains other than ASV made similar observations. For example, Wang et al. \cite{wang2019balanced} showed that data balancing may not be sufficient to address biases, and they attribute such behavior to bias amplification by models. Recently, in the field of ASR, Garnerin et al. \cite{garnerin2021investigating} showed that when training with balanced datasets, the actual speaker composition in the training data plays a key role in the biases observed in the system outputs.
We observe that using the proposed techniques to transform the x-vector speaker embeddings by including demographic information during training (UAI-AT and UAI-MTL) improves the fairness of systems considering the biases in FARs between the female and male population. The FaDR values ($\omega=1.0$) of the proposed methods (green and blue curves) remain close to 100\% at different values of the demographic-agnostic FAR. Therefore, in scenarios where we care about how many impostor verification pairs are incorrectly accepted by the ASV systems, the proposed embedding transformation techniques are beneficial in improving fairness with respect to gender.

We evaluate the FaDR at $\omega=0.5$ (denoting equal contribution of FAR and FRR), as shown in Figure \ref{fig:res_test10_0.5}, to consider the scenario where the discrepancy between genders in both the FAs and FRs of the system is of interest.
First, compared to the x-vector system trained on imbalanced data (orange curve), the system trained with data balanced with respect to the genders (red curve) performs better in terms of fairness across all operating points. This confirms the observation by Fenu et al. \cite{fenu2020improving}, that data balancing helps improve the fairness of speaker verification systems to some extent.
The proposed UAI-MTL and UAI-AT methods (green and blue curves) consistently perform better than the baselines in terms of fairness at all operating points (with the exception of UAI-MTL at FAR=1\%, where it is only slightly lower than the x-vector-B system). These results suggest that both adversarial and multi-task learning of speaker embeddings using gender labels can further improve the fairness of speaker representations compared to data balancing techniques.

An additional observation from the plots in Figure \ref{fig:res_test10} is that the benefits in terms of fairness compared to the baselines are more prominent at higher FARs. This is evident from the increasing difference between the FaDR values of the baseline x-vector-B and the proposed systems as the demographic-agnostic FAR increases. As we will see later in Section \ref{ssec:res_qualit}, this behavior can be explained by the distribution of the verification scores.
Also, FaDR only captures the absolute discrepancy in the performance between genders, but does not provide information about which particular demographic group is impacted. We discuss the performance of the systems separately for each gender group in \ref{app:bias_dir} to understand how the systems perform for each gender group separately.

\subsection{Fairness-Utility analysis}
\label{ssec:res_utility}
\begin{figure*}[t!]
\centering
\includegraphics[width=0.7\linewidth]{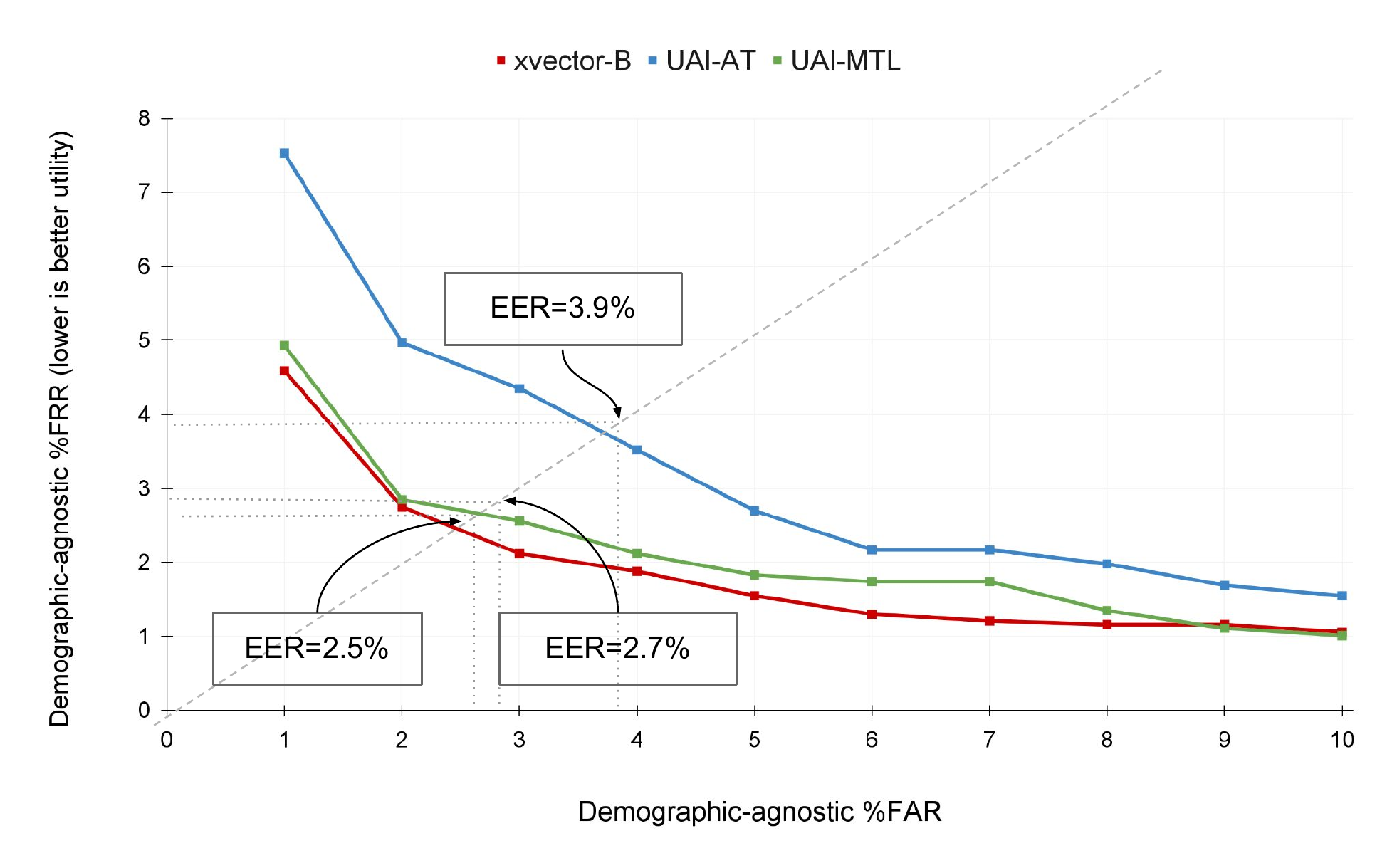}
\caption{Plot of demographic-agnostic \%FRR versus demographic-agnostic \%FAR showing the utility of the systems. Curves closer to the origin indicate better utility. Notice that the UAI-MTL system closely follows the baseline x-vector-B system at a range of operating conditions characterized by demographic-agnostic \%FAR. In contrast, the UAI-AT method reduces utility by increasing the \%FRR relative to the baseline x-vector-B system.}
\label{fig:res_utility}
\vspace{-0.15in}
\end{figure*}

Table \ref{tab:res_AUC_test} shows the area under the FaDR-FAR curve (auFaDR-FAR) along with the \%EER on the eval-test dataset. The auFaDR-FAR values are computed at $5$ different values of the error discrepancy weight, $\omega$. The auFaDR-FAR metric provides a quantitative summary of the fairness of the systems over several system operating points, with higher values denoting better fairness. The \%EER values are indicative of the speaker verification performance of the systems, providing an understanding of their utility, with lower values denoting better utility. In Figure \ref{fig:res_utility}, we show the FRR plotted against FAR using demographic-agnostic verification pairs. This describes the overall utility of the system, computed without considering the demographic attributes of the speakers. Curves closer to the origin denote better utility. In Table \ref{tab:res_Vox}, we report the results from the corresponding set of experiments on the voxceleb-H dataset, that help understand the generalizability of the proposed methods to more challenging in-the-wild recording conditions.
\begin{table}[t!]
\centering
\caption{auFaDR-FAR capturing fairness (binary gender groups) for $5$ different values of $\omega$, and \%EER capturing utility on \textbf{eval-test dataset}. Both the UAI-AT and UAI-MTL methods achieve similar auFaDR-FAR values, higher than the baseline x-vector-B for all values of $\omega$, with significant improvement from x-vector-B for $\omega=1.0$ (when discrepancy only in FAR between genders is considered) and $\omega=0.75,0.5$ (when discrepancy in FAR is weighted higher or equal to the discrepancy in FRR). UAI-MTL improves fairness while retaining utility (similar \%EER as x-vector-B), while UAI-AT achieves desired fairness at the cost of reduced utility. 
$\ast$ denotes significant improvement over the x-vector-B system (significance computed at level=0.01 using permutation test with $n=10000$ random permutations). The values in bold denote the highest fairness for each different value of $\omega$. The upper bound for auFaDR-FAR is 900, corresponding to perfect fairness (=100\% FaDR) at all values of demographic-agnostic FAR between 1\% and 10\% in Figure \ref{fig:res_test10}.}
\label{tab:res_AUC_test}
\resizebox{0.98\textwidth}{!}{
\begin{tabular}{c|c|cc|cc|cccc}
\toprule
  \multirow{2}{*}{Metric}   & \multirow{2}{*}{$\omega$}  &   \multicolumn{2}{c|}{\textbf{Baselines}} & \multicolumn{2}{c|}{\textbf{Proposed}}  & \multicolumn{4}{c}{\textbf{Ablations}} \\
                           &           & x-vector-U & x-vector-B & UAI-AT  & UAI-MTL & NLDR & UAI & {AT} & {MTL}  \\ \midrule
\multirow{5}{*}{auFaDR-FAR $\uparrow$} & $1.00$ &     842.3              & 864.9 & $\ 892.4^{\ast}$  & ${\ 892.5^{\ast}}$   & 840.2     &   863.9     & $\mathbf{\ 895.5^{\ast}}$ & 853.6   \\ 
& $0.75$ &        854.6             & 872.1 & ${\ 893.3^{\ast}}$  & $\ 892.9^{\ast}$ &  853.9   &  871.7  & $\mathbf{\ 894.9^{\ast}}$ & 864.6    \\
                           & $0.50$ &       866.8              & 879.2    & ${\ 894.3^{\ast}}$ & $\ 893.2^{\ast}$      & 867.6 &   879.5  & $\mathbf{\ 894.4^{\ast}}$ & 875.6   \\
& $0.25$ &         878.9            &  886.3  & $\mathbf{895.2}$  & 893.6  & 881.2 & 887.3    & 893.9 & 886.5  \\
                           & $0.00$ &       891.2              & 893.5    & $\mathbf{896.1}$  & 893.9  & 894.9 & 895.2    & 893.3 & 897.5  \\ \midrule
\%EER $\downarrow$   &  -         & 3.8                & 2.5 & $3.9$                & 2.7    &  2.5     &    ${2.4}$ & 2.8           & 2.7                   \\ \bottomrule
\end{tabular}
}
\end{table}

\begin{table}[ht]
\centering
\caption{auFaDR-FAR capturing fairness (binary gender groups) for $5$ different values of $\omega$, and \%EER capturing utility on \textbf{voxceleb-H dataset}. The UAI-MTL method achieves significantly higher auFaDR-FAR values than the baseline x-vector-B for all values of $\omega$. In contrast the UAI-AT method has a reduced fairness compared to the baseline. This suggests that the UAI-MTL technique to improve fairness generalizes across datasets, while the performance improvements from the UAI-AT technique seems to be inconclusive across datasets. Similar observations can be made with the AT technique, where performance gains observed in other datasets are not transferred to voxceleb-H. The values in bold denote the highest fairness for each different value of $\omega$. Omitting the results from x-vector-U, NLDR and UAI based on observations in Table \ref{tab:res_AUC_test}.} 
\label{tab:res_Vox}
\resizebox{0.75\textwidth}{!}{
\begin{tabular}{cc|c|cc|cc}
\toprule
                \multirow{2}{*}{Metric}   & \multirow{2}{*}{$\omega$}  &   \textbf{Baseline} & \multicolumn{2}{c|}{\textbf{Proposed}}  & \multicolumn{2}{c}{\textbf{Select ablations}} \\
                           &           & x-vector-B & UAI-AT  & UAI-MTL & AT & MTL  \\ \midrule
\multirow{5}{*}{auFaDR-FPR $\uparrow$} & $1.00$ &   863.9           &    848.4    & $\mathbf{872.2^{\ast}}$ & 859.9 & 820.6\\
& $0.75$ & 863.7 & 845.5 & $\mathbf{877.8^{\ast}}$ & 859.8 & 809.2 \\
                          & $0.50$ &  865.0            & 843.4 & $\mathbf{883.9^{\ast}}$ & 859.6 & 797.9 \\
& $0.25$ & 866.3 & 841.3 & $\mathbf{890.0^{\ast}}$ &  859.5 & 786.6 \\
                          & $0.00$ &    870.7         &    841.8  &  $\mathbf{898.4^{\ast}}$  &  859.3 & 775.3 \\ \midrule
\%EER $\downarrow$ & -               &  30.1             & 30.4  & 29.1 & 30.9 & 30.6  \\ \bottomrule
\end{tabular}
}
\end{table}
\textbf{Discussion}: The results in Table \ref{tab:res_AUC_test} show that the UAI-AT and UAI-MTL methods perform better than the x-vector baselines across all the values of $\omega$ examined. In particular, we found significant improvements using the proposed methods compared to the x-vector-B baseline at $\omega$ = 1.00, 0.75, 0.50. This confirms the statistical significance of the findings reported in the previous section. Additionally, we observe that the UAI-MTL method provides markedly better utility (as shown by the lower \%EER) than the UAI-AT system. This is also evident from Figure \ref{fig:res_utility}, where we observe that the UAI-MTL method (green curve) performs similar to the baseline x-vector-B speaker embeddings (red curve) in terms of speaker verification performance. Though the UAI-AT method performs similar to the UAI-MTL method in terms of fairness (auFaDR-FAR in Table \ref{tab:res_AUC_test}), it comes at the cost of degraded utility relative to the baseline x-vector-B speaker embeddings (shown by the shift of the blue curve away from the origin in Figure \ref{fig:res_utility}).
In summary, we find that the proposed multi-task method of transforming speaker embeddings can improve fairness to supplement data balancing techniques, while having minimal impact on utility (with statistically insignificant increase from 2.47 to 2.70). In contrast, the adversarial training method UAI-AT improves fairness at the cost of a significant increase in the \%EER (from 2.47 to 3.86). This suggests that multi-task learning using the UAI-MTL framework to transform speaker embeddings provides greater benefits than adversarial methods considering both improvement in the fairness of ASV systems and their impact on utility.

We observe from the ablation studies that the NLDR and UAI techniques to transform speaker embeddings are not effective at improving fairness. This shows that merely using speaker labels without the demographic information can not provide improvements in fairness over pre-trained speaker embeddings. This implies that the discriminator in Figure \ref{fig:unifai} is an indispensable module to mitigate biases present in existing speaker representations, as noted in previous work \cite{jaiswal2019unified}.
We also observe that MTL (without the UAI branch) is not effective in improving fairness. Even though AT shows some promise, we observe that the utility takes a hit (higher \%EER). Furthermore, we show in \ref{app:bias_weight} that adversarially trained methods (UAI-AT and AT) have greater variation in \%EER with respect to the contribution of the bias term on the training loss ($\delta$ in Equation \ref{eq:uai-param}). This makes it challenging to tune the bias weight. Also, as we discuss later using results on the voxceleb-H dataset, the AT and MTL techniques (without the UAI branch) do not generalize well to out-of-domain datasets. These experiments suggest that both the discriminator and the disentangler modules play an important role in developing fair speaker representations using the proposed methods.

We also note the relationship between the benefits of the proposed systems and the error discrepancy weight $\omega$ in Table \ref{tab:res_AUC_test}. We observe that as $\omega$ becomes smaller, the differences between the proposed techniques (UAI-AT,UAI-MTL) and the baseline x-vector-B system becomes smaller. This shows that, in applications with a greater emphasis on the discrepancy in the FAs of the system (higher values of $\omega$), the proposed techniques can be beneficial in improving the fairness of the baseline x-vector-B speaker embeddings. In applications where the emphasis is on the discrepancy in the FRs of the system (smaller values of $\omega$), the baseline x-vector-B embeddings are already fair, and the proposed techniques merely retain the fairness. We made a similar observation in Section \ref{ssec:res_fair} from Figure \ref{fig:res_test10}.


Results on the out-of-domain voxceleb-H test set are shown in Table \ref{tab:res_Vox}. We observe that the UAI-MTL technique to transform x-vector-B speaker embeddings attains the best performance in terms of fairness (highest auFaDR-FAR), and utility (lowest \%EER\footnote{The seemingly poor performance of utility of all the methods on the voxceleb-H dataset can be attributed to the utility of the x-vector speaker embeddings that we begin with. Here, our goal was to improve fairness of speaker embeddings, while retaining their utility.}). 
This suggests generalizability of multi-task training using the UAI framework when evaluated on a different dataset that is unseen during training.
We also observe that, performance improvements over the baselines using UAI-AT and AT techniques are inconsistent comparing with the results on the eval-test data shown in Table \ref{tab:res_AUC_test}, while UAI-MTL shows conclusive performance improvements even on out-of-domain data.
\begin{figure}[t!]
\centering
\begin{subfigure}[b]{.48\linewidth}
\includegraphics[width=\linewidth]{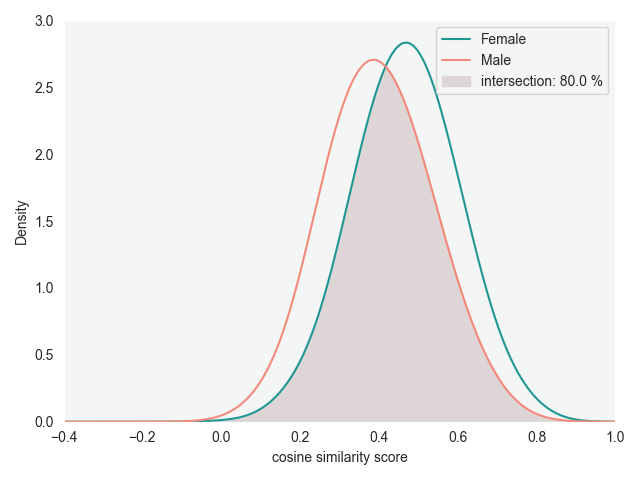}
\caption{x-vector-U}\label{fig:hist_xvec_Imbal}
\end{subfigure}
\begin{subfigure}[b]{.48\linewidth}
\includegraphics[width=\linewidth]{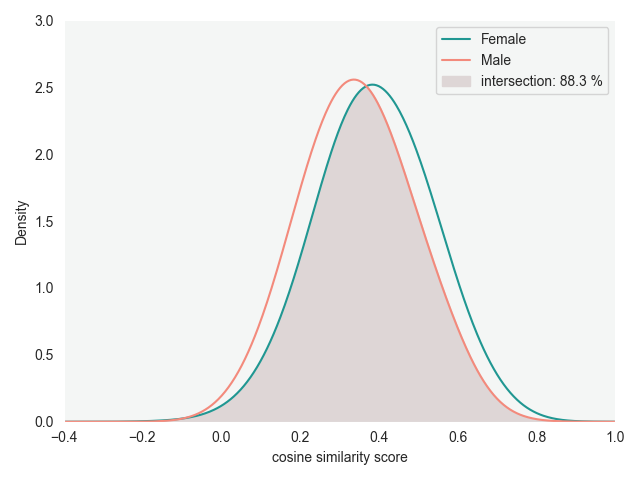}
\caption{x-vector-B}\label{fig:hist_xvec_bal}
\end{subfigure}

\begin{subfigure}[b]{.48\linewidth}
\includegraphics[width=\linewidth]{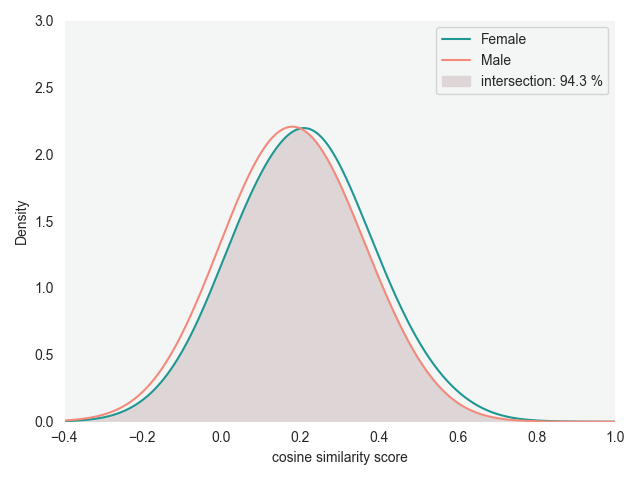}
\caption{UAI-AT}\label{fig:hist_uai_at}
\end{subfigure}
\begin{subfigure}[b]{.48\linewidth}
\includegraphics[width=\linewidth]{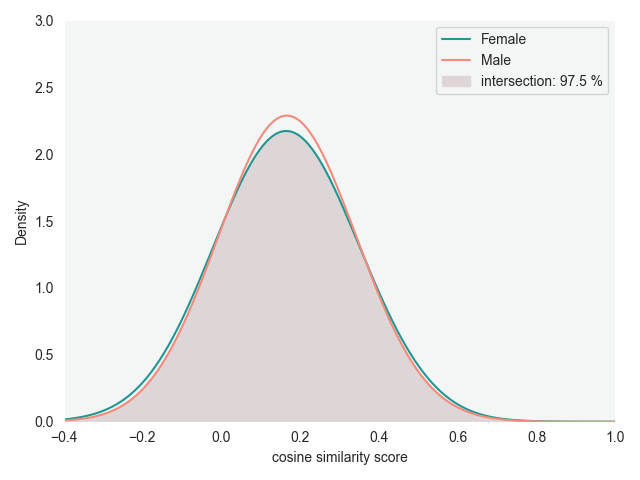}
\caption{UAI-MTL}\label{fig:hist_uai_mtl}
\end{subfigure}

\caption{Kernel density estimates of cosine similarity scores of \textbf{impostor pairs} (where test utterance belongs to speaker different from claimed identity) for the female and male demographic groups. Can be used to explain FAs. Both the x-vector baselines have the scores of the female population shifted compared to the scores of the male population, though training on balanced data (x-vector-B) seems to reduce the differences compared to x-vector-U. Transforming the x-vectors using UAI-AT and UAI-MTL techniques reduces differences between the scores of the female and male populations. Particularly, UAI-MTL produces scores with barely noticeable difference between the genders shown by the \%intersection in the scores between genders. This helps explain the observed improvement in fairness in Table \ref{tab:res_AUC_test} when considering the discrepancy in FAR between the female and male demographic groups ($\omega=1.0$).
}
\label{fig:kde}
\end{figure}
\subsection{Biases in verification scores}
\label{ssec:res_qualit}
\begin{figure}[t!]
\centering
\begin{subfigure}[b]{.48\linewidth}
\includegraphics[width=\linewidth]{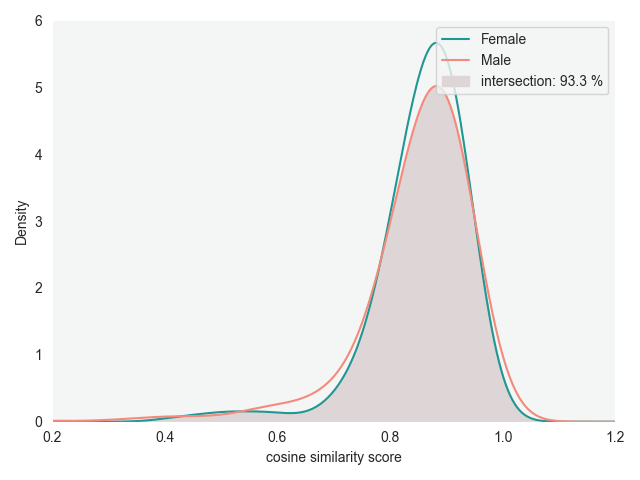}
\caption{x-vector-U}\label{fig:hist_xvec_Imbal_genuine}
\end{subfigure}
\begin{subfigure}[b]{.48\linewidth}
\includegraphics[width=\linewidth]{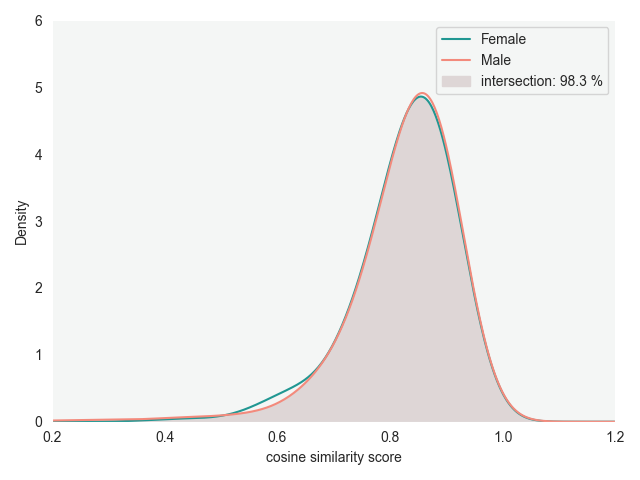}
\caption{x-vector-B}\label{fig:hist_xvec_bal_genuine}
\end{subfigure}

\begin{subfigure}[b]{.48\linewidth}
\includegraphics[width=\linewidth]{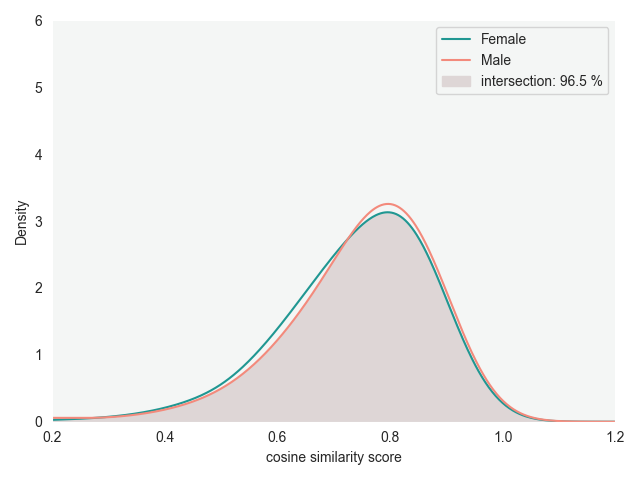}
\caption{UAI-AT}\label{fig:hist_uai_at_genuine}
\end{subfigure}
\begin{subfigure}[b]{.48\linewidth}
\includegraphics[width=\linewidth]{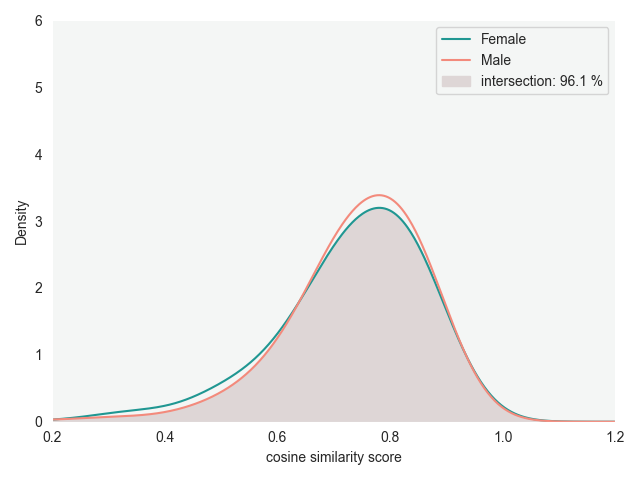}
\caption{UAI-MTL}\label{fig:hist_uai_mtl_genuine}
\end{subfigure}

\caption{Kernel density estimates of cosine similarity scores of \textbf{genuine pairs} (where test utterance belongs to same speaker as the claimed identity) for the female and male demographic groups. Can be used to explain FRs. Both the x-vector baselines have the scores of the female and male population overlapping with each other, indicating minimal bias between genders. This suggests that further embedding transformation techniques may not be necessary to improve the biases in the genuine verification scores between genders. This helps explain the high level of fairness observed in Table \ref{tab:res_AUC_test} when considering the discrepancy in FRR between the female and male demographic groups ($\omega=0.0$). It is worth noting that both the transformation techniques (UAI-AT and UAI-MTL) retain this overlap as shown by the \%intersection in the scores between genders.}
\label{fig:kde_target}
\end{figure}

Measures of fairness and utility are obtained after applying a threshold on the speaker verification scores as described in Section \ref{ssec:lit_asv}. We have quantitatively observed that the fairness of ASV systems can be improved through the UAI-AT and UAI-MTL techniques compared to the baseline x-vector systems. However, the primary source for lack of fair performance in ASV systems is the biases present in the speaker verification scores \cite{stoll2011finding}. In order to understand the biases present in the ASV systems, we perform a qualitative analysis of the verification scores similar to the work by Touissant et al. \cite{toussaint2021sveva}. In particular, we plot the kernel density estimate plots of the cosine similarity scores of the impostor verification pairs for the female and male populations in Figure \ref{fig:kde}, and those of the genuine pairs in Figure \ref{fig:kde_target}. The impostor verification scores determine the FARs, while the scores of the genuine verification pairs determine the FRRs of the systems.

\textbf{Discussion}: First, we notice from Figure \ref{fig:hist_xvec_Imbal} that there exists a skew in the cosine similarity scores between the female and male populations in the x-vector system trained on data imbalanced with respect to the genders. This points to the presence of biases, likely exacerbated by the training data imbalance. Particularly, we notice that the impostor scores for the female demographic population are higher than the scores of the male population, suggesting that at a given threshold, the proportion of FAs for the female population would be higher than for the male population. Such differences between the female and male impostor scores have been documented in prior literature \cite{marras2019adversarial}. Further, from Figure \ref{fig:hist_xvec_bal} it can be observed that training using data balanced with respect to the gender labels can mitigate the skew to some extent. 
Finally, we observe from Figures \ref{fig:hist_uai_at},\ref{fig:hist_uai_mtl} that both the adversarial and multi-task learning techniques can further reduce the skew between the female and male verification scores. In particular, the UAI-MTL method produces almost overlapping score distributions for the female and male populations. This suggests that transformation of the speaker embeddings using gender information in a multi-task fashion using the UAI-MTL framework can help mitigate the biases in the impostor verification scores between genders. Subsequent application of a threshold on these scores would therefore produce similar rates of FAs for the female and male populations, as we have seen in Section \ref{ssec:res_fair}.

From Figures \ref{fig:hist_xvec_Imbal_genuine} and \ref{fig:hist_xvec_bal_genuine}, we notice that the scores of the genuine pairs between the female and male demographic groups are mostly overlapping. This implies that at any given threshold, we would not observe much difference between the proportions of FRs of the demographic groups. This is consistent with the quantitative analysis shown earlier in Table \ref{tab:res_AUC_test}, where we found high values of the auFaDR-FAR metric for the x-vector systems computed at smaller values of $\omega$ (corresponding to greater emphasis on differences in FRR between genders).
We observe from the figures \ref{fig:hist_uai_at_genuine} and \ref{fig:hist_uai_mtl_genuine} that embedding transformation using the proposed methods retains the unbiased nature of the genuine verification scores obtained from the pre-trained embeddings. 
In summary, we show that the proposed methods improve or retain the fairness depending on the target use-case. In scenarios where the rates of false accepts are an important consideration, the proposed UAI-AT and UAI-MTL methods are able to reduce the biases present in existing speaker representations. When the false rejects are more important, our methods preserve the fairness of existing speaker representations.

\section{Conclusions and future directions}
\label{sec:conclusion}
We presented adversarial and multi-task learning strategies to improve the fairness of extant speaker embeddings with respect to demographic attributes of the speakers. 
In the adversarial setting, the demographic attribute labels were used to learn speaker embeddings devoid of the demographic information. 
In the multi-task approach, the goal was to learn demographic-aware speaker embeddings, where the demographic information is explicitly infused into the embeddings. In particular, we adopted the unsupervised adversarial invariance (UAI) framework \cite{jaiswal2019unified} to investigate whether adversarial or multi-task training is better suited for reducing the biases with respect to binary gender groups in speaker embeddings used in ASV systems. 
We used the recently proposed fairness discrepancy rate metric \cite{de2020fairness} to evaluate the fairness of the systems at various operating points. We observed that data balancing,  a commonly used strategy to improve fairness, mitigates the biases to some extent. However, its fairness depends on the operating point of interest (whether it is a low FAR or low FRR operating region). Therefore it is important to consider the specific application -- and the corresponding desired operating region -- of the ASV systems when evaluating fairness. 
For applications strictly focused on the differences between  genders in their FRRs, existing x-vector speaker embeddings (either trained on balanced or imbalanced data) performed well by having very minimal biases, and the speaker embeddings transformed using the proposed methods retained this desirable property. However, as we move toward applications focused on the differences between the genders in their FARs, the x-vector speaker embeddings showed biases between the genders. In this scenario, the proposed adversarial and multi-task training strategies were able to mitigate these biases by a significant margin. Furthermore, we showed qualitative evidence that the proposed methods were able to effectively reduce the biases in the verification score distributions between the female and male populations.
In addition, we showed that it is critical to jointly consider aspects of both fairness and utility in selecting embedding transformation techniques. We found that the adversarial and multi-task training strategies showed similar performance on fairness metrics. However, while multi-task training to transform the x-vector speaker embeddings had very little impact on the utility, the adversarial training strategy significantly degraded the utility.

We explored several aspects of fairness and utility of ASV systems in this work. However, we believe that there are still open questions that require further investigation.
We have limited our analyses to gender as the demographic factor of interest in our investigations. However, considering other demographic attributes (including intersectional demographics) is important \cite{9101635}. For example, systems that are not biased with respect to the gender alone could be biased when a different demographic factor (e.g. age) is considered as an intersecting attribute.
Also, we trained our models using the MCV corpus, and analyzed the biases in these systems using the MCV and Voxceleb corpora. However, such datasets could be prone to systemic censoring \cite{kallus2018residual}. For example, the MCV corpus may not be sufficiently representative of the different demographic groups and their intersectional attributes, because the data was collected only from users with access to a microphone and internet connection.
Similarly the Voxceleb corpus consists of speech samples only from celebrities. 
A more inclusive adoption of such technologies requires careful consideration of these various aspects, which we hope to address in future research. Finally, as mentioned before, we adopted notions of biases which belong to the category of group fairness. However, individual fairness, which is an alternate way of evaluating biases, can also provide interesting insights into how these systems behave. 
\pagebreak
\appendix

\section{Effect of bias weight}
\label{app:bias_weight}
We trained several models by varying the weight parameter $\delta$ in Equation \ref{eq:uai-param}. This parameter allowed us to control the influence of the discriminator loss on the overall optimization. As described in Section \ref{sec:experiment}, we fixed the values for the weights of the predictor, decoder and disentangler modules based on preliminary experiments to $\alpha=100$, $\beta=5$ and $\gamma=100$ respectively. Therefore, by varying $\delta$ we studied the isolated effect of the discriminator loss on the training objective.

\begin{table}[!t]
\centering
\caption{Classification results on embed-val dataset and verification results on eval-dev dataset for different bias weights ($\delta$ in Equation \ref{eq:uai-param}). The majority class random chance accuracy for bias labels in the embed-val data was $70$\%}
\label{tab:res_val}
\resizebox{0.65\textwidth}{!}{
\begin{tabular}{cc|cc|c|c}
\toprule
                                                              & bias weight & \begin{tabular}[c]{@{}c@{}}\%acc. \\ (predictor)\end{tabular} & \begin{tabular}[c]{@{}c@{}}\%acc. \\ (bias)\end{tabular} & \%EER & auFaDR                       \\ \midrule
\begin{tabular}[c]{@{}c@{}}xvector-U\end{tabular} & -           & -                                                             & -                                                        & 2.66 & 871.09 \\
\begin{tabular}[c]{@{}c@{}}xvector-B \end{tabular}   & -           & -                                                             & -                                                        & 2.36  & 884.57 \\ \midrule
\multirow{7}{*}{UAI-AT}                                       & 10           & 96.99                                                         & 78.24                                                    & 2.81 & 893.40                      \\
                                                              & 30          & 96.91                                                         & 70.55                                                    & 3.42 & 892.17                     \\
                                                              & 50          & 96.58                                                         & 75.92                                                    & 4.95 & 893.03                     \\
                                                              & 70          & 96.76                                                         & 80.35                                                    & 3.12 & \textbf{896.38}            \\
                                                              & 100          & 96.6                                                          & 82.03                                                    & 4.11 & 890.21                     \\
                                                              & 150          & 96.1                                                          & 77.70                                                    & 4.58 & 888.12                     \\
                                                              & 200         & 95.32                                                         & 72.80                                                    & 10.26 & 893.69                     \\ \midrule
\multirow{7}{*}{UAI-MTL}                                      & 10           & 97.04                                                         & 97.03                                                    & 2.45  & \textbf{896.72}            \\
                                                              & 30          & 96.99                                                         & 97.98                                                    & 2.66 & 885.45                     \\
                                                              & 50          & 96.96                                                         & 98.52                                                    & 2.70 & 886.52                     \\
                                                              & 70          & 97.01                                                         & 98.73                                                    & 3.06 & 858.99                     \\
                                                              & 100          & 96.96                                                         & 98.86                                                    & 2.66 & 848.88                     \\
                                                              & 150          & 96.94                                                         & 99.02                                                    & 2.98 & 851.31                     \\
                                                              & 200         & 96.91                                                         & 99.03                                                    & 3.19 & 852.62                     \\ \midrule
\multirow{5}{*}{AT} & 10 & 96.72 & 69.91 & 3.00 & 879.68\\ 
& 30 & 96.71 & 76.04 & 3.40 & 882.82\\ & 50 & 96.65 & 71.91 & 3.14 & \textbf{897.04} \\ & 70 & 96.24 & 59.37 & 9.78 & 884.28\\ & 100 & 95.91 & 78.10 & 8.11 & 884.06\\ \midrule
\multirow{5}{*}{MTL} &  10 & 96.68 & 97.19 & 2.47 & 861.27 \\ & 30 & 96.70 & 98.33 & 2.52 & \textbf{876.25} \\ & 50 & 96.79 & 98.71 & 2.77 & 858.71 \\ & 70 & 96.75 & 98.91 & 2.99 & 852.92 \\ & 100 & 93.73 & 98.96 & 2.75 & 870.16 \\
\bottomrule
\end{tabular}
}
\end{table}

\textbf{Discussion}: The second pair of columns in Table \ref{tab:res_val} shows the speaker classification accuracy of the predictor and gender classification accuracy of the discriminator on the embed-val dataset. Clearly, the UAI-AT method is able to reduce the gender classification accuracy to close to majority class chance performance ($70$\%). This shows that the technique is able to successfully reduce the amount of gender information in the speaker embeddings. On the other hand, owing to its multi-task training setup, the UAI-MTL method retains gender information in the speaker embeddings. This is evident from the high gender classification accuracy of the discriminator ($>97\%$).

Verification results on the eval-dev dataset are shown in the third set of columns in Table \ref{tab:res_val}. We notice that compared to the UAI-AT models, the UAI-MTL models provide better verification performance as shown by the \%EER in all settings. In addition, across different training configurations (characterized by the bias weights), the UAI-MTL method has a smaller variation in \%EER (min:$2.45$, max:$3.19$) when compared with the UAI-AT method (min:$2.81$, max:$10.26$). This provides further evidence of the negative impact on the utility of adversarial training when compared with multi-task learning. It validates the findings from prior research that have shown the instability of adversarial training \cite{sadeghi2021adversarial}. We find similar trends in models trained without the UAI branch. Specifically, we observe that the MTL methods have a smaller variation in \%EER (min:$2.47$, max:$2.99$) when compared to the AT methods (min:$3.00$, max:$9.78$).
Finally, for all the methods, we choose the optimal bias weight $\delta$ based on the best auFaDR-FAR value (in bold). This model was used for the evaluations on the eval-test dataset that were described in Section \ref{sec:results}.

\section{Direction of bias}
\label{app:bias_dir}
\begin{figure}[t!]
\centering
\begin{subfigure}[b]{.325\linewidth}
\includegraphics[width=\linewidth]{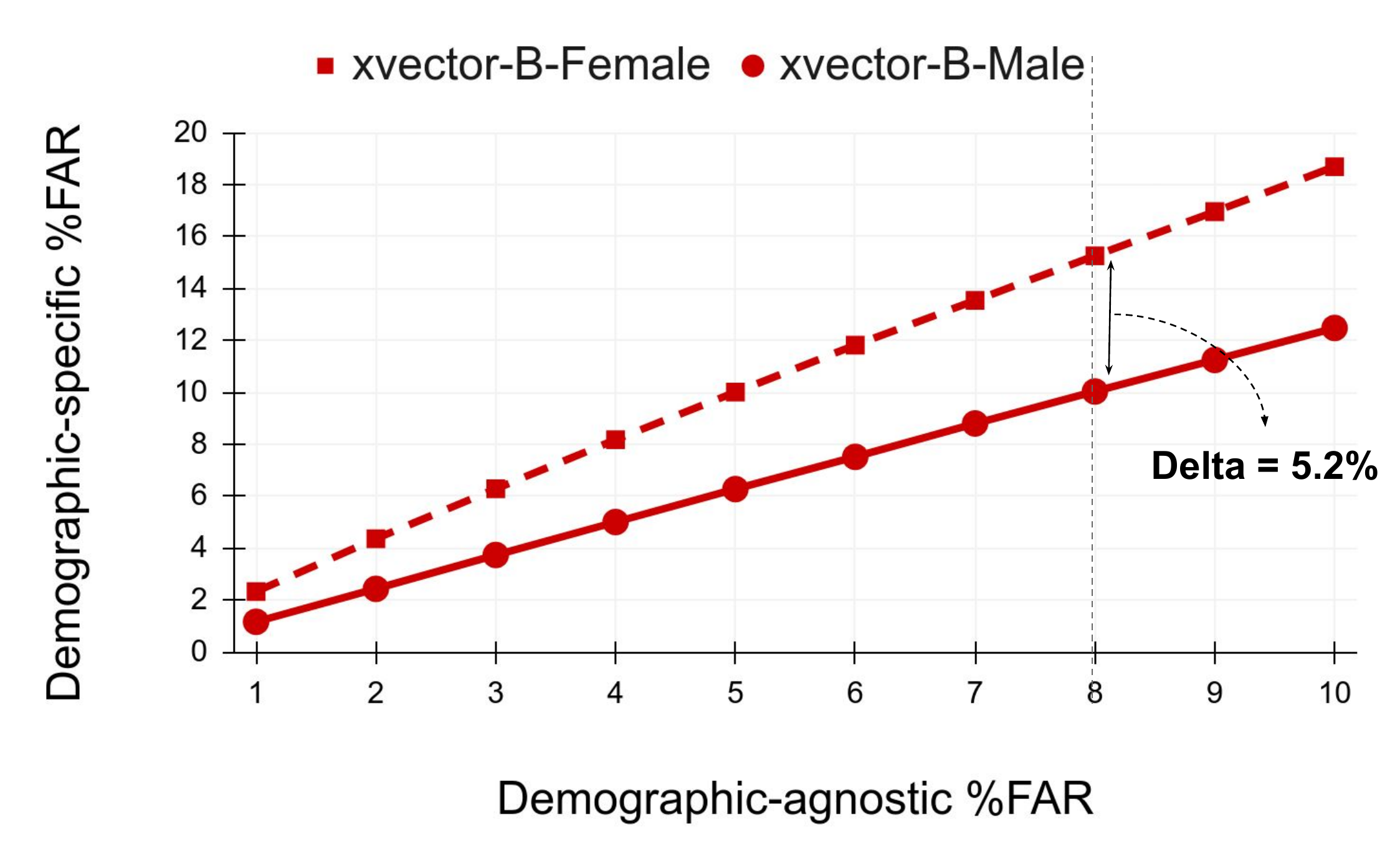}
\caption{Demographic-specific FAR : xvector-B}\label{fig:res_dem_spec_FPR_xvec}
\end{subfigure}
\begin{subfigure}[b]{.325\linewidth}
\includegraphics[width=\linewidth]{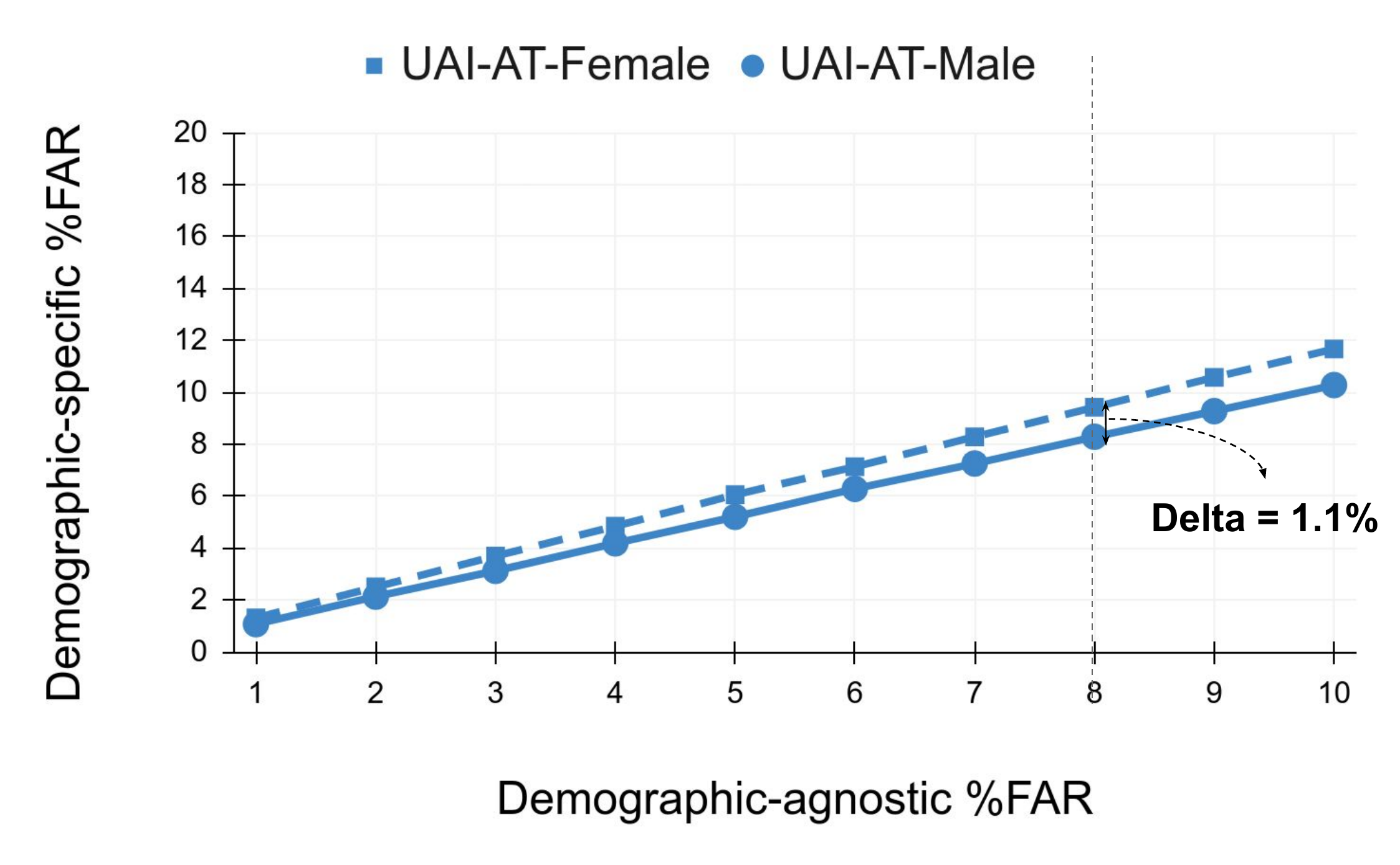}
\caption{Demographic-specific FAR : UAI-AT}\label{fig:res_dem_spec_FPR_uai_at}
\end{subfigure}
\begin{subfigure}[b]{.325\linewidth}
\includegraphics[width=\linewidth]{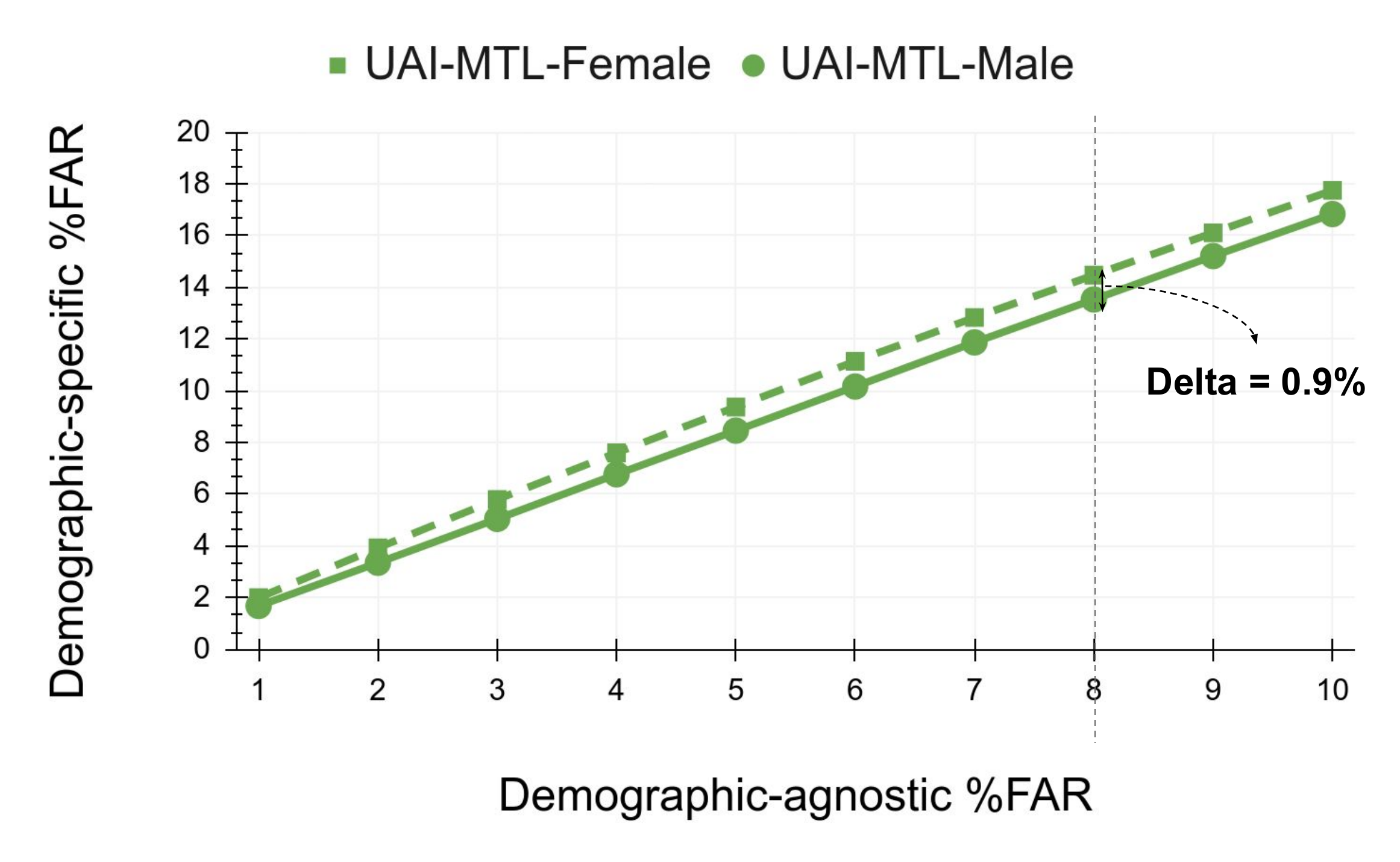}
\caption{Demographic-specific FAR : UAI-MTL}\label{fig:res_dem_spec_FPR_uai_mtl}
\end{subfigure}

\caption{Demographic-specific FAR for the baseline and proposed systems. Here, unlike FaDR which considers absolute discrepancy in the FAR, we look at the individual FARs of the female and male populations. This gives an indication whether a particular demographic groups is particularly impacted due to biases in the systems. Notice how, compared to the x-vector systems, both the proposed methods reduce the difference in the FARs between the female and male populations. However, the UAI-AT method achieves this by reducing the EER for both groups, while the UAI-MTL method achieves this by increasing the error rate for the male population making it closer to the female population.}
\label{fig:res_dem_specific_FPR}
\end{figure}

\begin{figure}[t!]
\centering
\begin{subfigure}[b]{.325\linewidth}
\includegraphics[width=\linewidth]{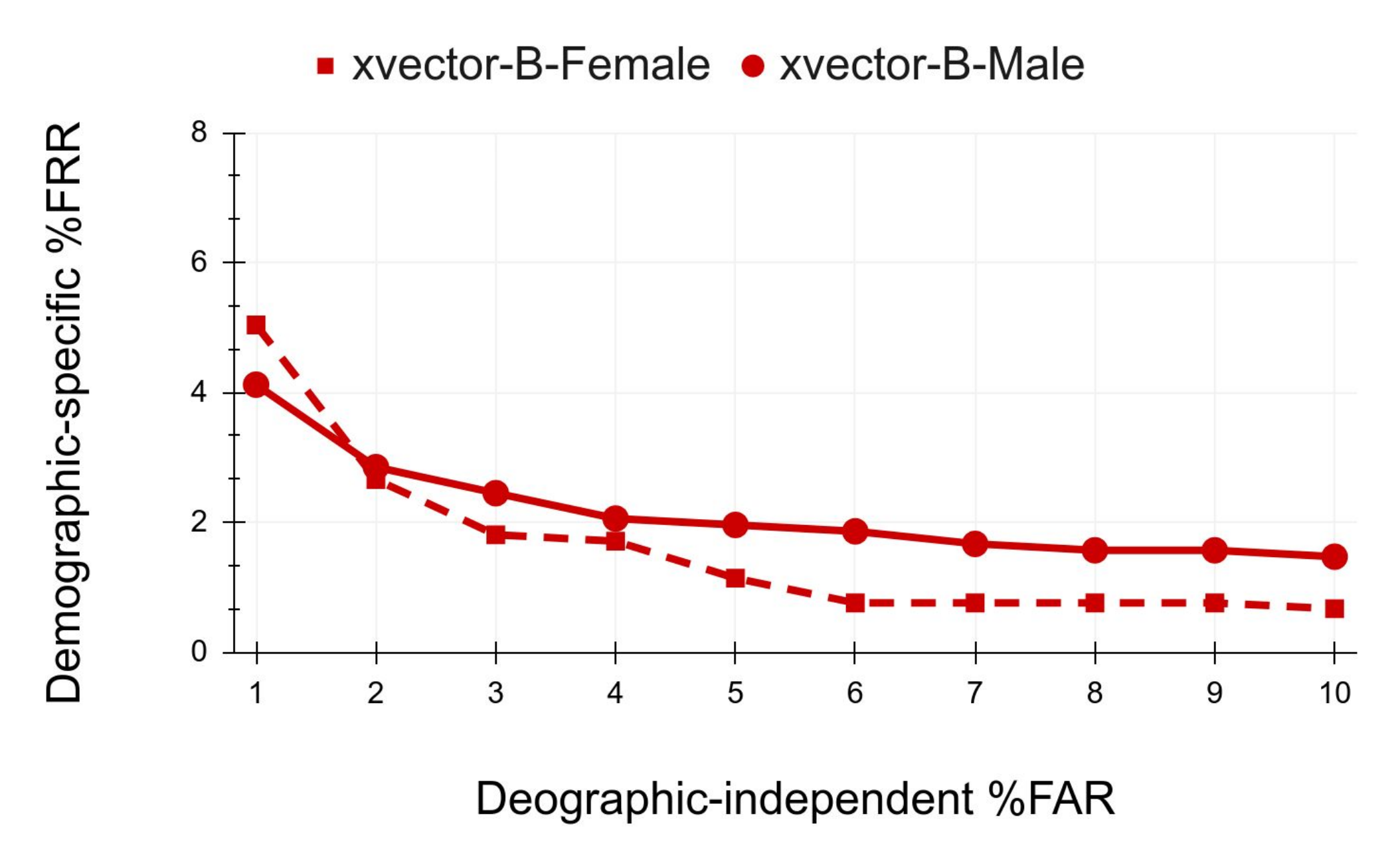}
\caption{Demographic-specific FRR : xvector-B}\label{fig:res_dem_spec_FNR_xvec}
\end{subfigure}
\begin{subfigure}[b]{.325\linewidth}
\includegraphics[width=\linewidth]{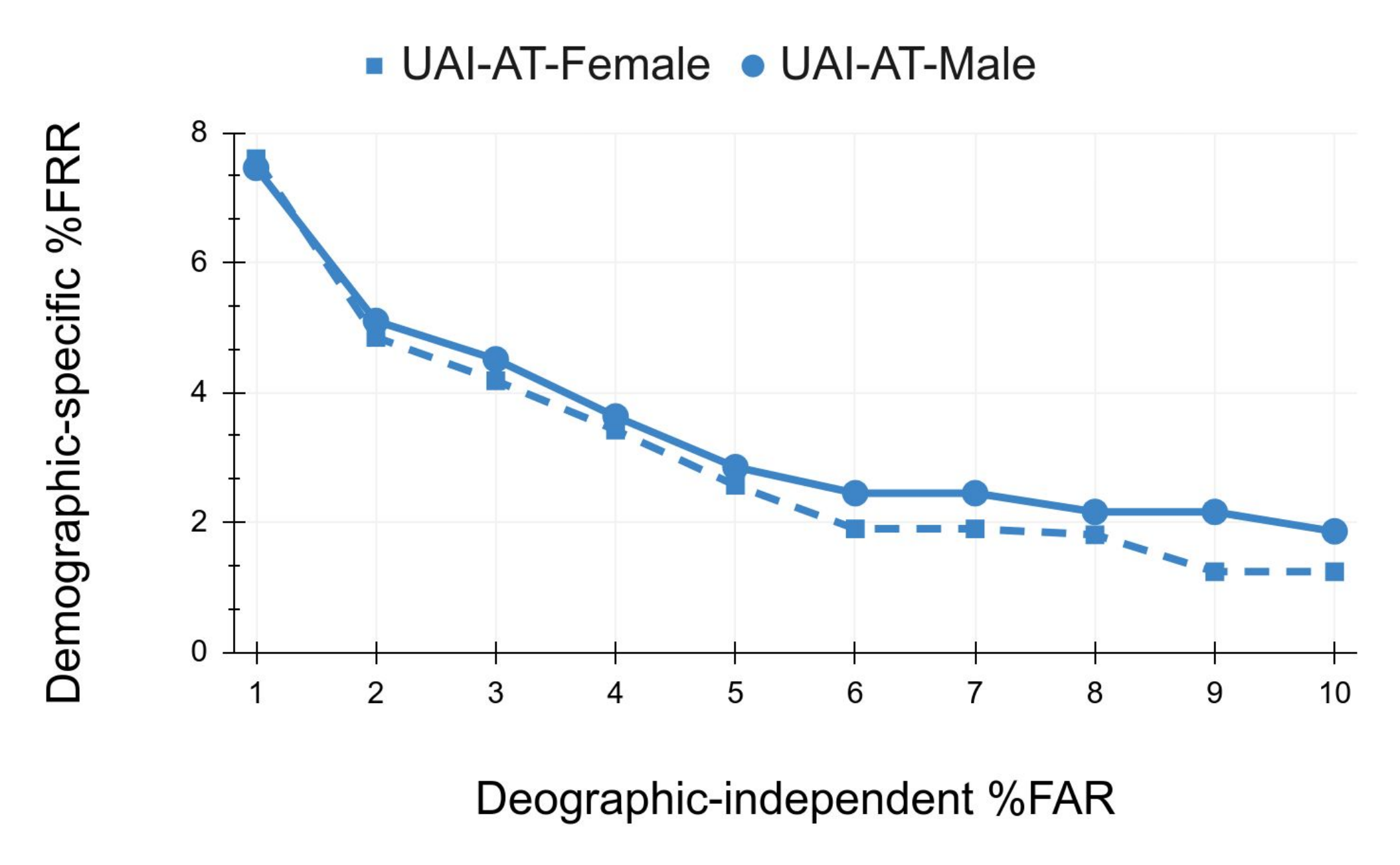}
\caption{Demographic-specific FRR : UAI-AT}\label{fig:res_dem_spec_FNR_uai_at}
\end{subfigure}
\begin{subfigure}[b]{.325\linewidth}
\includegraphics[width=\linewidth]{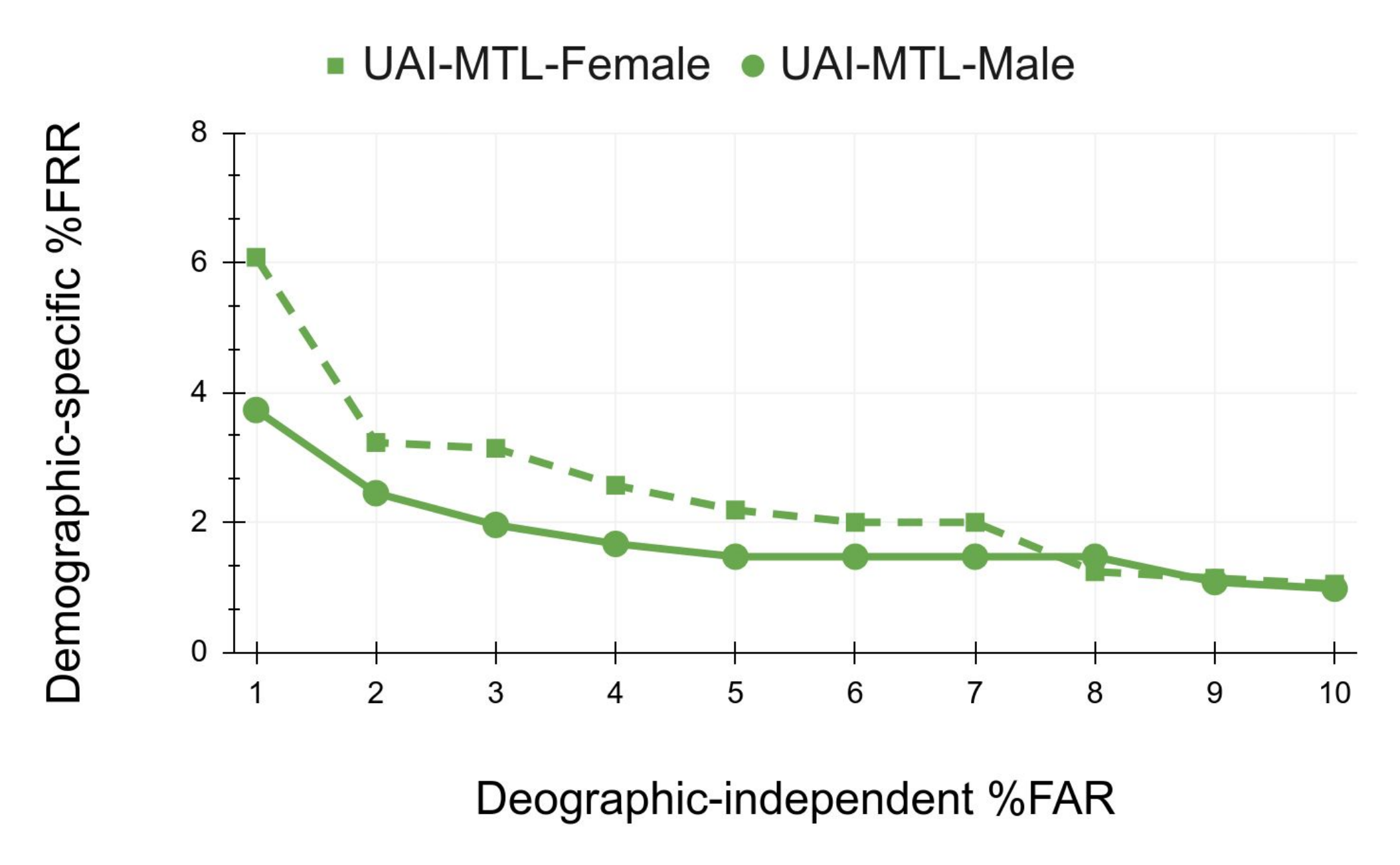}
\caption{Demographic-specific FRR : UAI-MTL}\label{fig:res_dem_spec_FNR_uai_mtl}
\end{subfigure}

\caption{Demographic-specific FRR for the baseline and proposed systems. Here, unlike FaDR which considers absolute discrepancy in the FRR, we look at the individual FRRs of the female and male populations. This gives an indication whether a particular demographic groups is particularly impacted due to biases in the systems. Notice how the baseline x-vector method already shows very little differences in the FRRs between the female and male populations. Further, the proposed techniques retain this small differences.}
\label{fig:res_dem_specific_FNR}
\end{figure}

In Section \ref{sec:results}, we reported the results using FaDR metric, which considers the absolute difference between the FARs and FRRs of the female and male demographic groups. It does not provide a sense of the direction of bias. Previous studies have shown that ASV systems are prone to higher error rates for the female population than the male population \cite{george2015cosine, fenu2020exploring}. In a similar vein, we wanted to investigate if there is a systematic bias against a particular gender. In particular, we wanted to check if the ASV systems consistently underperform for a particular demographic group when compared with a different demographic group. We report the individual FARs~(\ref{fig:res_dem_specific_FPR}) and FRRs(Figure \ref{fig:res_dem_specific_FNR}) of the female and male populations at varying thresholds characterized by demographic-agnostic \%FARs.

\textbf{Discussion}: From Figure \ref{fig:res_dem_specific_FPR}, we observe that the baseline x-vector system is highly biased against the female demographic groups considering FARs. This is evident from the gap between the curves for the male (solid blue) and female (dotted blue) populations. Furthermore, the gap increases at higher values of demographic-agnostic \%FAR. On the other hand, both the proposed UAI-AT and UAI-MTL methods reduce the gap in \%FAR between the female and male populations. However, they show noticeably different behavior. The UAI-MTL reduces the \%FAR of the female population (dotted red) compared to the x-vector baseline, while simultaneously increasing the \%FAR of the male population (solid red), bringing them closer to each other. On the other hand, the UAI-AT method substantially reduces the \%FAR on the female population (dotted green), while also reducing the \%FAR on the male population (solid green) by a small extent. At first glance, this seems to suggest that UAI-AT is a better technique since it improves the performance of both demographic groups with respect to \%FAR. However, as we discussed in Section \ref{ssec:res_utility}, considering the \%FRR of the systems, UAI-AT method degrades the performance, thereby affecting the overall utility of the ASV system.

In Figure \ref{fig:res_dem_specific_FNR}, we report the \%FRR for the female and male populations. Notice the difference in the scale of y-axis compared to Figure \ref{fig:res_dem_specific_FPR}. Here, we observe that there is not much difference between the \%FRRs of the different demographic groups even with the baseline x-vector system. Furthermore, we observe that the UAI-MTL method to transform x-vectors does not have a substantial impact on the performance compared with x-vectors. The UAI-AT technique of transforming x-vectors increases \%FRR for both the female and male populations to some extent.

\pagebreak



\bibliographystyle{elsarticle-num}
\bibliography{refs}
\end{spacing}




\biboptions{sort&compress}
\end{document}